\documentclass[prd,onecolumn,aps,nofootinbib,eqsecnum,notitlepage,nobibnotes,superscriptaddress,preprintnumbers]{revtex4-1}

\usepackage[utf8]{inputenc}
\usepackage{color}
\usepackage{amsfonts}
\usepackage{graphicx}
\usepackage[dvipsnames]{xcolor}
\definecolor{refs}{RGB}{245,156,74}
 \usepackage[colorlinks=true,hyperfootnotes=true,citecolor=cyan]{hyperref}
 
\usepackage{dsfont} 

\pdfoutput=1

\usepackage{amsmath,amsthm,amssymb}
\usepackage{bm}

\usepackage{dcolumn}

\allowdisplaybreaks[1]

\begin{document}
\newcommand{\newc}{\newcommand}

\newc{\rk}[1]{{\color{red} #1}}
\newc{\ben}{\begin{eqnarray}}
\newc{\een}{\end{eqnarray}}
\newc{\be}{\begin{equation}}
\newc{\ee}{\end{equation}}
\newc{\ba}{\begin{eqnarray}}
\newc{\ea}{\end{eqnarray}}
\newc{\D}{\partial}
\newc{\rH}{{\rm H}}
\newc{\cH}{{\mathcal H}}
\newc{\dphi}{\delta\phi}
\newc{\pa}{\partial}
\newc{\tp}{\dot{\phi}}
\newc{\ttp}{\ddot{\phi}}
\newc{\drhoc}{\delta\rho_c}
\newc{\aB}{\alpha_{\rm B}}
\newc{\aK}{\alpha_{\rm K}}
\newc{\aM}{\alpha_{\rm M}}
\newc{\bn}{\beta_{n_c}}
\newc{\bK}{\beta_{\rm K}}
\newc{\delc}{\delta_{c{\rm N}}}
\newc{\eH}{\epsilon_{\rm H}}
\newc{\ep}{\epsilon_{\phi}}

\newc{\cs}{c_{\rm s}}
\newc{\dd}{{\rm d}}

\newcommand{\jose}[1]{{\color{blue}{#1}}}

\title{Velocity-dependent interacting dark energy and dark matter \\
with a Lagrangian description of perfect fluids}

\author{Jose Beltr\'an Jim\'enez}
\email{jose.beltran@usal.es}
\affiliation{Departamento~de~F{\'i}sica~Fundamental~and~IUFFyM,~Universidad~de~Salamanca,~E-37008~Salamanca,~Spain.}
\author{Dario Bettoni}
\email{bettoni@usal.es}
\affiliation{Departamento~de~F{\'i}sica~Fundamental~and~IUFFyM,~Universidad~de~Salamanca,~E-37008~Salamanca,~Spain.}
\author{David Figueruelo}
\email{davidfiguer@usal.es}
\affiliation{Departamento~de~F{\'i}sica~Fundamental~and~IUFFyM,~Universidad~de~Salamanca,~E-37008~Salamanca,~Spain.}
\author{Florencia A. Teppa Pannia}
\email{f.a.teppa.pannia@usal.es}
\affiliation{Departamento~de~F{\'i}sica~Fundamental~and~IUFFyM,~Universidad~de~Salamanca,~E-37008~Salamanca,~Spain.}
\author{Shinji Tsujikawa}
\email{tsujikawa@waseda.jp}
\affiliation{Department of Physics, Waseda University, 3-4-1 Okubo, Shinjuku, Tokyo 169-8555, Japan.}

\begin{abstract}
We consider a cosmological scenario where the dark sector is described by two perfect fluids that interact through a velocity-dependent coupling. This coupling gives rise to an interaction in the dark sector driven by the relative velocity of the components, thus making the background evolution oblivious to the interaction and only the perturbed Euler equations are affected at first order. We obtain the equations governing this system with the Schutz-Sorkin Lagrangian formulation for perfect fluids
and derive the corresponding stability conditions to avoid ghosts and Laplacian instabilities. As a particular example, we study a model 
where dark energy behaves as a radiation fluid at high redshift while it effectively becomes a cosmological constant in the late Universe. 
Within this scenario, we show that the interaction of both dark components leads to a suppression of the dark matter clustering at late times. 
We also argue the possibility that this suppression of clustering together with the additional dark radiation at early times can simultaneously alleviate the $\sigma_8$ and $H_0$ tensions.
\end{abstract}

\date{\today}
\maketitle
\section{Introduction}
\label{introsec}

In the last decades, Cosmology has turned from being mostly speculative, where precise data was barely available to test the different cosmological models, to a data-driven science. 
This is attributed to the great efforts made to gather high-precision 
data from the Cosmic Microwave Background (CMB) \cite{Spergel:2003cb,Ade:2013zuv}, 
type Ia supernovae \cite{Riess:1998cb,Perlmutter:1998np}, 
galaxy surveys \cite{Eisenstein:2005su,Tegmark:2006az,Blake:2011rj}, weak lensing \cite{Hildebrandt:2016iqg,Abbott:2017wau}, etc. 
All these data have allowed to establish a standard model for cosmology, dubbed the $\Lambda$CDM \cite{Peebles:1984ge,Peebles:1982ff}, 
where the present-day Universe is mostly dominated by Cold Dark Matter (CDM) and Dark Energy (DE) in the form of a cosmological constant $\Lambda$. Despite some theoretical challenges posed 
by this model \cite{Weinberg:1988cp,Martin:2012bt}, at a phenomenological level it has shown a fairly good agreement 
with most of data and hence the $\Lambda$CDM has been 
regarded as the standard cosmological paradigm.

However, as the amount of cosmological information as well as 
its precision increases, some discrepancies among different 
observations start to arise between high- and 
low-redshifts such as the tensions of today's Hubble constant $H_0=100h$~km\,s$^{-1}$\,Mpc$^{-1}$ \cite{Riess:2016jrr,Aghanim:2018eyx,Verde:2019ivm,Riess:2019cxk,
Wong:2019kwg,Reid:2019tiq} 
and the amplitude of matter perturbations 
$\sigma_8$ within the comoving $8h^{-1}$\,Mpc scale \cite{Macaulay:2013swa,Nesseris:2017vor,Joudaki:2017zdt}. 
Although such tensions may be due to unknown systematics, 
they could also be signalling the presence of new physics 
beyond the $\Lambda$CDM model.
To address the problem of $H_0$ tension, there have been 
a number of theoretical 
attempts \cite{Karwal:2016vyq,Poulin:2018cxd,Agrawal:2019lmo} 
to modify the early 
cosmological dynamics by taking into account a scalar 
field which initially behaves as a cosmological constant 
and subsequently decays faster than non-relativistic matter. 
The presence of early DE reduces the sound 
horizon around the CMB decoupling epoch, so the value of 
$H_0$ can be larger than that in the $\Lambda$CDM.
However, it was recently shown that the early DE 
does not completely alleviate the $H_0$ tension 
by including the large-scale structure data 
besides the CMB data in the 
analysis \cite{Hill:2020osr,Ivanov:2020ril,DAmico:2020ods}.

The other possible way to ease the $H_0$ tension is to consider 
the late-time DE with a phantom equation of state 
($w_d<-1$) \cite{DiValentino:2016hlg,Vagnozzi:2019ezj}. 
While the standard canonical scalar field 
like quintessence cannot realize $w_d<-1$ without 
the appearance of ghosts, the scalar or vector field 
with derivative interactions or non-minimal couplings to 
gravity \cite{Horndeski:1974wa,Deffayet:2011gz,Kobayashi:2011nu,Heisenberg:2014rta,Tasinato:2014eka,Jimenez:2016isa} 
gives rise to a phantom equation of state without 
theoretical inconsistencies \cite{Tsujikawa:2010zza}. 
Indeed, there are models of late-time cosmic acceleration 
in the framework of scalar-tensor or vector-tensor theories 
which can reduce the $H_0$ 
tension \cite{Peirone:2019aua,DeFelice:2016yws,deFelice:2017paw,DeFelice:2020sdq,Heisenberg:2020xak}. 
On the other hand, the modified gravity models with 
the speed of gravity equivalent to that of light
usually lead to the cosmic growth rate larger than 
that in the $\Lambda$CDM 
model \cite{DeFelice:2011hq,Tsujikawa:2015mga,Amendola:2017orw,
Kase:2018aps}, 
so it is hard to address the problem of $\sigma_8$ tension 
without any direct interaction between DE and CDM.

If the DE field is coupled to CDM through an energy 
transfer, the CDM perturbation usually grows faster 
in comparison to the 
$\Lambda$CDM model \cite{Amendola:2003wa,Tsujikawa:2007gd,Ade:2015rim}. 
If there is a momentum exchange between DE and CDM, 
the growth of CDM perturbations can slow down due to 
the suppression of the CDM velocity potential. 
For a canonical scalar field $\phi$ (quintessence) 
coupled to the CDM four-velocity $u_c^{\mu}$ 
through the scalar product 
$Z=u_c^{\mu} \partial_{\mu} \phi$, 
the weak cosmic growth can be realized by 
the momentum 
transfer \cite{Pourtsidou:2013nha,Boehmer:2015sha,Skordis:2015yra,Koivisto:2015qua,Pourtsidou:2016ico,Dutta:2017kch,Linton:2017ged,Kase:2019veo,Kase:2019mox,Chamings:2019kcl}.
Indeed, the likelihood analysis of Ref.~\cite{Pourtsidou:2016ico} 
for a concrete quintessence model with the interacting Lagrangian
$f \propto Z^2$ alleviates the $\sigma_8$ tension.
In Refs.~\cite{Amendola:2020ldb,Kase:2020hst}, it was shown that the suppression of the cosmic growth rate induced by the momentum transfer is generic even in more general scalar-tensor theories and in 
the presence of the energy transfer. 
This property also persists in vector-tensor theories 
with the vector field $A_{\mu}$ coupled to the CDM 
velocity in the form 
$u_c^{\mu}A_{\mu}$ \cite{DeFelice:2020icf}.

There are also interacting models where both DE and CDM are 
dealt as perfect fluids. The difference from quintessence 
is that the DE fluid can cluster, depending on its sound 
speed $c_{d}$ \cite{Erickson:2001bq,Bean:2003fb}. 
Moreover, unlike the cosmological constant, the energy 
density of the DE fluid can give rise to an additional 
contribution to the Hubble expansion rate 
at early times. 
Provided that the DE density is transiently important 
around radiation-matter equality, there is a possibility 
that the $H_0$ tension can be eased by the early 
DE fluid \cite{Lin:2019qug,Lin:2020jcb}. 
We note that this DE fluid is also different from the
scalar-field early DE followed by the oscillation
around its potential 
minimum \cite{Karwal:2016vyq,Poulin:2018cxd,Agrawal:2019lmo}, 
in that the latter has
the time-averaged values of $w_{d}$ and $c_{d}$ 
over oscillations.

In Refs.~\cite{Asghari:2019qld,Jimenez:2020ysu} the authors 
studied fluid DE models coupled to the CDM or baryon fluid, 
with the momentum exchange weighed by the difference 
between four velocities. In these works the starting point 
is not the covariant action of interacting fluids, but 
a covariant modification of the continuity equations of DE and CDM (or baryon) in terms of the relative 4-velocities of DE and the matter components.
Since the new term depends on the relative velocities, only the momentum conservation is modified so the background cosmological dynamics is not affected, but 
it leads to the suppression for the growth of 
matter perturbations at late times due to the dragging produced by the DE pressure on the matter components. 
These dark fluid models significantly improve 
the $\sigma_8$ tension in comparison to the $\Lambda$CDM.  

In this paper, we provide a Lagrangian formulation of 
the dark fluids interacting  through the momentum 
transfer. We employ the Schutz-Sorkin 
action \cite{Schutz:1977df,Brown:1992kc,DeFelice:2009bx} to describe 
both DE and CDM perfect fluids and consider the interacting 
Lagrangian of the form $f(Z)$, where $f$ is a function 
of the product $Z=u_{c}^{\mu}u_{d \mu}$ between CDM and 
DE four velocities. Unlike the phenomenological approaches 
taken in Refs.~\cite{Dalal:2001dt,Chimento:2003iea,Wang:2005jx,
Wei:2006ut,Amendola:2006dg,Guo:2007zk,Valiviita:2008iv,Salvatelli:2014zta,Kumar:2016zpg,DiValentino:2017iww,Yang:2018euj,Pan:2019gop,
DiValentino:2019jae}, the background and perturbation 
equations of motion unambiguously follow from the fully 
covariant action.
The perturbation equations 
are found to be different from those in Refs.~\cite{Asghari:2019qld,Jimenez:2020ysu}, 
but they share the common property that the momentum 
exchange is  determined by the relative velocities. A difference however arises since the scenario considered here also features a dependence on the relative acceleration that is absent in Refs.~\cite{Asghari:2019qld,Jimenez:2020ysu}. This represents a distinctive property of this model.
We particularise the general developed framework to a model where DE behaves 
as a dark radiation at early times and approaches a cosmological
constant at late times. In this model, we show that the growth 
rate of matter perturbations is suppressed by the momentum transfer, 
thus alleviating the $\sigma_8$ tension. 
Moreover, this model can potentially reduce the $H_0$ tension thanks to 
the early-time modification, but we leave the detailed 
likelihood analysis with recent observational data 
for a future work.

\section{Lagrangian description of coupled DE and DM}
\label{eomsec}

In this section, we introduce interacting theories of DE 
and CDM with a momentum exchange through their four velocities.
We consider a scenario where both DE and 
CDM are described by perfect fluids. 
This means that there exists a comoving frame 
in which they appear as isotropic and they are fully described 
by their densities and pressures. 
In principle, the comoving frames of both perfect fluids do not need to coincide 
with each other and the mixture can indeed behave as an effective non-perfect fluid where the momentum density and anisotropic stresses arise from the non-comoving state of both fluids. However, we will consider that the comoving frames coincide on sufficiently large scales as to comply with the cosmological principle dictating that our Universe is isotropic on such scales\footnote{Cosmological models with non-comoving fluids have been explored in e.g., Refs.~\cite{Maroto:2005kc,BeltranJimenez:2007rsj,Jimenez:2008vs,Harko:2013wsa,Cembranos:2019plq,Garcia-Garcia:2016dcw}. It would be interesting to extend our analysis to those scenarios.}. 

In the following, we will study an interaction between the fluids that is governed 
by their velocities. 
If $u^\mu_c$ and $u^\mu_d$ are the 4-velocities of the comoving frames of CDM and DE, respectively, the interaction must be a function of the only scalar that we can construct, which is given by
\be
Z\equiv g_{\mu \nu} u_c^{\mu} u_d^{\nu}\,,
\label{Zdef}
\ee
where $g_{\mu \nu}$ is the metric tensor.
Notice that this is the leading interaction at lowest order in derivatives. Couplings involving the four-accelerations of the fluids will be suppressed by some scale that also determines the scale at which additional (unstable) modes come in. 
Of course, the perfect fluid (or even the fluid) approximation might breakdown 
at a much lower scale where viscosity and anisotropic stresses become relevant. 
We will neglect all such deviations from perfection as well as 
the higher-derivatives operators.  

The system of the two dark fluids interacting via the coupling 
in Eq.~\eqref{Zdef}, including the gravitational sector, can then be described by the following action
\be
{\cal S} =
\frac{M_{\rm pl}^2}{2}\int {\rm d}^4 x \sqrt{-g}\,R
-\sum_{I=c,d,b,r}\int {\rm d}^{4}x 
\Big[\sqrt{-g}\,\rho_I(n_I)+ J_I^{\mu} \partial_{\mu} \ell_I \Big]
+\int {\rm d}^4x\sqrt{-g}\,f(Z)\,.
\label{action}
\ee
The first term is the usual Einstein-Hilbert action of General Relativity 
where $g$ is the determinant of $g_{\mu\nu}$, 
$M_{\rm pl}$ is the reduced Planck mass, 
and $R$ is the Ricci scalar.
The second integral in Eq.~(\ref{action}), which is known as a Schutz-Sorkin 
action \cite{Schutz:1977df,Brown:1992kc,DeFelice:2009bx}, describes the perfect fluids 
of CDM, DE, baryons, and radiation, labeled by $c,d,b,r$, 
respectively. \footnote{An alternative formalism to describe the dynamics of the scenario under consideration would be the effective field theory of perfect fluids applied to the case of several interacting components as done in e.g. \cite{Ballesteros:2013nwa}.} 
The energy density $\rho_I$ depends on each fluid number density 
$n_I$, where $n_I$ is related to the current vector field 
$J_I^{\mu}$ in the action (\ref{action}) as
\be
n_I=\sqrt{\frac{g_{\mu \nu} J_I^{\mu} J_I^{\nu}}{g}}\,.
\label{ndef}
\ee
The relation between $J_I^{\mu}$ and the four velocity 
$u_I^{\mu}$ is given by 
\be
J_I^{\mu}=n_I \sqrt{-g}\,u_I^{\mu}\,,
\label{Jmu}
\ee
which guarantees 
$g_{\mu \nu} u_I^{\mu}u_I^{\nu}=-1$ from Eq.~(\ref{ndef}).
The scalar quantity $\ell_I$ in the Schutz-Sorkin action 
is a Lagrange multiplier, 
with the notation of the partial derivative 
$\partial_{\mu} \ell_I \equiv \partial \ell_I/\partial x^{\mu}$ 
with respect to a coordinate $x^{\mu}$.
Finally, the last term in the action \eqref{action}, 
which depends on the arbitrary function $f(Z)$, 
represents the velocity-dependent coupling 
mediating a momentum exchange between CDM and DE. 
The quantity $Z$, defined in Eq.~(\ref{Zdef}), 
is expressed as 
\be
Z=-\frac{g_{\mu \nu} J_c^{\mu}J_d^{\nu}}{g\,n_c n_d}\,.
\ee
We assume that baryons and radiation are coupled to 
neither CDM nor DE. 

\subsection{Covariant equations of motion}

Having the full action for the system, we can proceed to 
obtain the corresponding covariant equations of motion.
Varying the action (\ref{action}) with respect to $\ell_I$, 
it follows that   
\be
\partial_{\mu}J_I^{\mu}=0\qquad
({\rm for}~I=c,d, b,r)\,,
\label{Jmucon}
\ee
which shows that the current $J_I^{\mu}$ is conserved.
Since $\partial_{\mu}(\sqrt{-g}\,u_I^{\mu})
=\sqrt{-g}\,\nabla_{\mu}u_I^{\mu}$, where $\nabla_{\mu}$ is 
the covariant derivative operator, Eq.~(\ref{Jmucon}) 
translates to $u_I^{\mu}\partial_{\mu}n_I+n_I\nabla_{\mu}u_I^{\mu}=0$. 
The energy density $\rho_I$ depends on $n_I$ alone, 
so there is the relation  
$\rho_{I,n_I}u_I^{\mu}\partial_{\mu}n_I=u_I^{\mu}\partial_{\mu}\rho_I$, 
where $\rho_{I,n_I} \equiv \partial \rho_I/\partial n_I$. 
Introducing the pressure of each fluid, 
\be
P_I=n_I\rho_{I,n_I}-\rho_I\,, 
\label{Pdef}
\ee
Eq.~(\ref{Jmucon}) can be expressed in the form, 
\be
u_I^{\mu}\partial_{\mu}\rho_I
+(\rho_I+P_I)\nabla_{\mu}u_I^{\mu}=0\,.
\label{umu}
\ee
This is the continuity equation for the  
energy-momentum tensor of each fluid.

To vary the action (\ref{action}) with respect to $J_I^{\mu}$, 
we exploit the following properties, 
\be
\frac{\partial n_I}{\partial J_I^{\mu}}=\frac{J_{I \mu}}{n_I g}\,,\qquad
\frac{\partial Z}{\partial J_c^{\mu}}=-\frac{1}{n_c g} 
\left( \frac{J_{d \mu}}{n_d}+\frac{Z J_{c \mu}}{n_c} \right)\,,\qquad
\frac{\partial Z}{\partial J_d^{\mu}}=-\frac{1}{n_d g} 
\left( \frac{J_{c \mu}}{n_c}+\frac{Z J_{d \mu}}{n_d} \right)\,.
\ee
Then, we find the following relations
\ba
\partial_{\mu} \ell_{c}
&=& \rho_{c,n_c} u_{c{\mu}}+\frac{f_{,Z}}{n_c}
\left( u_{d\mu}+Zu_{c\mu} \right) \,,
\label{lc} \\
\partial_{\mu} \ell_{d}
&=& \rho_{d,n_d} u_{d{\mu}}+\frac{f_{,Z}}{n_d}
\left( u_{c\mu}+Zu_{d\mu} \right) \,,
\label{ld} \\
\partial_{\mu} \ell_{I}
&=& \rho_{I,n_I} u_{I{\mu}} \qquad({\rm for}~I=b,r)\,,
\label{lbr} 
\ea
which are used to eliminate the Lagrange multipliers $\ell_I$ from 
the covariant equations of motion derived below.

We express the action (\ref{action}) in the form 
${\cal S}=\int {\rm d}^4 x\,(L_g+L_m)$, where 
\be
L_g=\sqrt{-g}\frac{M_{\rm pl}^2}{2}R\,,\qquad
L_m=-\sum_{I=c,d,b,r}\left[\sqrt{-g}\,\rho_I(n_I)
+J_I^{\mu} \partial_{\mu} \ell_I \right]
+\sqrt{-g}\,f(Z)\,.
\ee
Varying the Einstein-Hilbert Lagrangian $L_g$ 
with respect to $g^{\mu \nu}$, we have
\be
\frac{2}{\sqrt{-g}} \frac{\delta L_g}{\delta g^{\mu \nu}}
=M_{\rm pl}^2 G_{\mu \nu}\,,
\ee
where $G_{\mu \nu}$ is the Einstein tensor. 
For the variation of $L_m$ with respect to $g^{\mu \nu}$, 
we exploit the following relations
\be
\frac{\delta \sqrt{-g}}{\delta g^{\mu \nu}} = -\frac{1}{2} \sqrt{-g}\,
g_{\mu \nu}\,,
\qquad
\frac{\delta n_I}{\delta g^{\mu \nu}} = \frac{n_I}{2} \left( g_{\mu \nu} 
-u_{I\mu} u_{I\nu} \right)\,,\qquad
\frac{\delta Z}{\delta g^{\mu \nu}} =\frac{Z}{2} 
\left( u_{c\mu} u_{c \nu}
+u_{d\mu} u_{d \nu} \right)
+u_{c \mu} u_{d \nu}\,.
\ee
Then, it follows that 
\be
-\frac{2}{\sqrt{-g}} \frac{\delta L_m}{\delta g^{\mu \nu}}
=\sum_{I=c,d,b,r}T^{(I)}_{\mu\nu}
+T^{({\rm int})}_{\mu\nu}\,,
\label{Lfva}
\ee
where 
\ba
T^{(I)}_{\mu\nu}&=&(\rho_I+P_I)u_{I{\mu}}u_{I{\nu}}+P_Ig_{\mu\nu}\,,
\label{TI}\\
T^{({\rm int})}_{\mu\nu}&=&
f g_{\mu\nu}+f_{,Z} \left( Z u_{c\mu} u_{c \nu}
+Z u_{d\mu} u_{d \nu}+2 u_{d \mu} u_{c \nu} \right)\,.
\label{Tmunuin}
\ea
Then, the gravitational equations of motion are given by  
\be
M_{\rm pl}^2G_{\mu\nu}
=\sum_{I=c,d,b,r}T^{(I)}_{\mu\nu}
+T^{({\rm int})}_{\mu\nu}\,.
\label{ein}
\ee
Taking the covariant derivative of Eq.~(\ref{ein}), 
we obtain
\be
\sum_{I=c,d,b,r} \nabla^{\mu}T^{(I)}_{\mu\nu}
+\nabla^{\mu}T^{({\rm int})}_{\mu\nu}=0\,.
\label{conco}
\ee
On using Eq.~(\ref{umu}), the perfect-fluid 
energy-momentum tensor $T^{(I)}_{\mu\nu}$ obeys
\be
u_I^{\nu} \nabla^{\mu}T^{(I)}_{\mu\nu}
=-\left[ u_I^{\mu}\partial_{\mu}\rho_I
+(\rho_I+P_I)\nabla_{\mu}u_I^{\mu} \right]=0\,,
\label{ucov}
\ee
which is equivalent to the continuity Eq.~(\ref{Jmucon}).
If the four-velocities of CDM, DE, baryons, and radiation 
are identical to each other (which is the case for the 
isotropic and homogeneous cosmological background), 
then the continuity equation 
$u^{\nu} \nabla^{\mu}T^{(I)}_{\mu\nu}=0$ holds 
for each fluid or a single fluid with the four-velocity 
$u^{\nu}$.
In this case, Eq.~(\ref{conco}) gives 
$u^{\nu} \nabla^{\mu}T^{({\rm int})}_{\mu\nu}=0$. 
This property does not hold for the four-velocity of 
each fluid different from each other 
(as in the case of a perturbed spacetime).

\subsection{Background equations of motion} 
\label{bacsec}
As explained above, the velocity-dependent coupling that we consider is chosen so that the background evolution is not modified. This property follows from the fact that all the cosmological fluids are assumed to share a common rest frame on sufficiently large scales that we can associate to the CMB rest frame where the metric is given by the spatially flat 
Friedmann-Lema\^{i}tre-Robertson-Walker (FLRW) line element
\be
{\rm d}s^2=-{\rm d}t^2
+a^2(t) \delta_{ij} {\rm d}x^i {\rm d}x^j\,,
\label{BGmet}
\ee
with $a(t)$ the scale factor.
Each perfect fluid in this rest frame has the four-velocity
$u_I^{\mu}=(1,0,0,0)$, with $I=c,d,b,r$. 
Since $J_I^0=n_I a^3$ from Eq.~(\ref{Jmu}), 
the constraint Eq.~(\ref{Jmucon}) gives  
\be
J_I^0 \equiv {\cal N}_I=n_I a^3={\rm constant}\,.
\label{ncon}
\ee
This means that the particle number ${\cal N}_I$ of 
each fluid is conserved. 
{}From Eq.~(\ref{umu}) it follows that 
\be
\dot{\rho}_I+3H \left( \rho_I+P_I \right)=0\,,\qquad {\rm for} 
\quad I=c,d,b,r,
\label{coneq}
\ee
where a dot represents the derivative 
with respect to $t$, and $H=\dot{a}/a$ 
is the Hubble-Lema\^{i}tre expansion rate.
The continuity Eq.~(\ref{coneq}) is equivalent to 
the particle number conservation Eq.~(\ref{ncon}).

The (00) and $(ii)$ components of the gravitational Eq.~(\ref{ein})
lead to the following background equations
\ba
3M_{\rm pl}^2 H^2&=&\sum_{I=c,d,b,r}\rho_I-f\,,\label{back1}\\
M_{\rm pl}^2 \left( 2\dot{H}+3H^2 \right)
&=&-\sum_{I=c,d,b,r}P_I-f\,.
\label{back2}
\ea
Since $Z=-1$ on the background (\ref{BGmet}), the $f_{,Z}$-dependent terms 
in Eq.~(\ref{Tmunuin}) do not contribute to the background Eqs.~(\ref{back1}) 
and (\ref{back2}). However, the coupling $f$ itself, which is constant for the background configuration, 
affects the background dynamics.
One can absorb this cosmological constant term into the definitions 
of $\rho_d$ and $P_d$, such that 
\be
\hat{\rho}_d=\rho_d-f\,,\qquad
\hat{P}_d=P_d+f\,.
\label{hatrhoP}
\ee
These effective dark energy density and pressure obey
\be
\dot{\hat{\rho}}_d+3H \left( \hat{\rho}_d
+\hat{P}_d \right)=0\,.
\ee
Then, the right hand-sides of Eqs.~(\ref{back1}) 
and (\ref{back2}) are expressed as $\hat{\rho}_d+\rho_c+\rho_b+\rho_r$ and 
$-\hat{P}_d-P_c-P_b-P_r$, respectively.

\section{Cosmological perturbations}
\label{persec}

In this section, we derive all the linear perturbation equations of motion 
on the flat FLRW background without choosing a particular gauge. 
The line element containing four scalar perturbations 
$\alpha,\chi,\zeta$ and $E$ is given by  
\be
{\rm d}s^2=-(1+2\alpha) {\rm d}t^2
+2 \partial_i \chi {\rm d}t {\rm d}x^i
+a^2(t) \left[ (1+2\zeta) \delta_{ij}
+2\partial_i \partial_j E \right] {\rm d}x^i {\rm d}x^j\,,
\label{permet}
\ee
where the perturbations depend on both cosmic time 
$t$ and spatial coordinates $x^i$. 
The temporal and spatial components of $J_I^{\mu}$ 
are decomposed as
\be
J_I^{0}={\cal N}_I+\delta J_I\,,\qquad 
J_I^{i}=\frac{1}{a^2(t)}\delta^{ik}\partial_k\delta j_I\,, 
\label{JI}
\ee
where ${\cal N}_I$ is the background conserved
number of each particle, and $\delta J_I$ and $\delta j_I$ 
correspond to scalar perturbations. 
Substituting Eq.~(\ref{JI}) into Eq.~(\ref{ndef}), the perturbation 
of particle number density $n_I$, which is expanded 
up to second order, yields
\be
\delta n_I=\frac{{\cal N}_I}{a^3} 
\left[ \frac{\delta\rho_I}{\rho_{I}+P_I}
-\frac{\delta\rho_I}{\rho_{I}+P_I} \left( 3\zeta+\partial^2 E\right)
-\frac{(\partial \delta j_I+{\cal N}_I \partial\chi)^2}{2 {\cal N}_I^2 a^2}
-\frac{1}{2}(\zeta+\partial^2E)(3\zeta-\partial^2E)\right]\,,
\label{deln}
\ee
where $\delta\rho_I$ is the density perturbation defined by 
\be
\delta\rho_I=\frac{\rho_{I}+P_I}{{\cal N}_I} \left[ \delta J_I-{\cal N}_I 
\left( 3\zeta+\partial^2 E \right) \right]\,.
\label{drhoI}
\ee
At linear order, $\delta\rho_I$ is related to $\delta n_I$ according to 
$\delta\rho_I=\rho_{I,n_I}\delta n_I$. 
The four velocity $u_{I \mu}=J_{I \mu}/(n_I \sqrt{-g})$, 
which is expanded up to linear order, is given by 
\be
u_{I0}=-1-\alpha\,,\qquad
u_{Ii}=-\partial_i v_I\,, 
\label{uI}
\ee
where $v_I$ corresponds to the velocity potential 
related to $\delta j_I$ and $\chi$, as
\be
v_I=-\frac{\delta j_I}{{\cal N}_I}-\chi\,.
\label{vI}
\ee
{}From Eqs.~(\ref{drhoI}) and (\ref{vI}), one can express 
$\delta J_I$ and $\delta j_I$ in terms of $\delta \rho_I$, 
$v_I$, and metric perturbations. 

The energy density $\rho_I$, which depends on $n_I$ alone, 
is expanded as
\be
\rho_I (n_I)=\rho_I+\left( \rho_I+P_I \right) 
\frac{\delta n_I}{n_I}+\frac{1}{2} \left( \rho_I+P_I \right) 
c_I^2 \left( \frac{\delta n_I}{n_I} \right)^2
+{\cal O} (\varepsilon^3)\,,
\ee
where $c_I^2$ is the adiabatic sound speed squared  
defined by 
\be
c_I^2=\frac{n_I \rho_{I,n_I n_I}}{\rho_{I,n_I}}
=\frac{\dot{P}_I}{\dot{\rho}_I}\,.
\label{cI}
\ee
We also introduce the fluid equation of state parameter 
\be
w_I=\frac{P_I}{\rho_I}\,,
\ee
whose time derivative is related to the 
adiabatic sound speed by
\be
\dot{w}_I=3H(1+w_I)\left(w_I-c_I^2\right).
\ee
This relation recovers the well-known fact that 
a perfect fluid with constant equation of state has $w_I=c_I^2$.

By using Eq.~(\ref{uI}) with the background value $Z=-1$, 
the spatial component of Eq.~(\ref{lc}), 
up to  linear order in perturbations, reads
\be
\partial_i \ell_c = -\rho_{c,n_c} \partial_i v_c
-\frac{f_{,Z}}{n_c} (\partial_i v_d-\partial_i v_c)\,, 
\label{ellc2}
\ee
where $\rho_{c,n_c}$, $n_c$, and $f_{,Z}$ need to be
evaluated on the background. 
The integration of Eq.~(\ref{ellc2}) with respect to 
$x^i$ gives rise to a time-dependent term ${\cal A}(t)$ as 
a global-in-space mode.
Since $\dot{\ell}_c=-\rho_{c,n_c}$ 
on the background, we have ${\cal A}(t)=-\int^t \rho_{c,n_c}(\tilde{t}) 
{\rm d} \tilde{t}$ and hence
\be
\ell_c=-\int^t \rho_{c,n_c}(\tilde{t}) {\rm d}\tilde{t}
-\rho_{c,n_c} v_c-\frac{f_{,Z}}{n_c} (v_d-v_c)\,.
\ee
This relation will be used to eliminate the Lagrange multiplier $\ell_c$
from the action (\ref{action}). 
Similarly from Eqs.~(\ref{ld}) and (\ref{lbr}), we obtain
\ba
\ell_d &=& -\int^t \rho_{d,n_d}(\tilde{t}) {\rm d}\tilde{t}
-\rho_{d,n_d} v_d-\frac{f_{,Z}}{n_d} (v_c-v_d)\,,\\
\ell_I &=& -\int^t \rho_{I,n_I}(\tilde{t}) {\rm d}\tilde{t}
-\rho_{I,n_I} v_I \qquad({\rm for}~I=b,r)\,.
\ea
The coupling $f(Z)$ is expanded as 
\be
f(Z)=f+f_{,Z} \delta Z\,,
\ee
where 
\be
\delta Z=-\frac{1}{2a^2} \left( \partial_i v_d
-\partial_i v_c \right)^2\,.
\ee
Since $\delta Z$ is of second order in perturbations, we do not need to 
expand $f(Z)$ up to the order of $f_{,ZZ} \delta Z^2/2$.

\subsection{Perturbation equations}

Now we are ready for expanding the action (\ref{action}) up to 
quadratic order in scalar perturbations. 
After the necessary integrations by parts, the second-order action is 
expressed in the form 
\be
{\cal S}^{(2)}=\int {\rm d}t\,{\rm d}^3x 
\left( L_g+L_m+L_{\rm int} \right)\,,
\label{S2}
\ee
where
\ba
L_g
&=& \frac{aM_{\rm pl}^2}{2} \left[ 2\{ 3a^2 H (2\dot{\zeta}+3H \zeta)
-2H \partial^2 \chi-2\partial^2 \zeta \} \alpha-3a^2 (2\dot{\zeta}^2+3H^2 \alpha^2)
+2 (\partial_i \zeta)^2+3H^2 (\partial_i \chi)^2+4\dot{\zeta} \partial^2 \chi \right]
\nonumber \\
& &+a^3 M_{\rm pl}^2 \left[ 2\ddot{\zeta}+2H (3\dot{\zeta}-\dot{\alpha}) 
\right]\partial^2 E+\frac{a^3 M_{\rm pl}^2}{2} \left( 2\dot{H}+3H^2 \right)
\left[ 3 \zeta^2+\partial^2 E (2\zeta-2\alpha-\partial^2 E) \right]\,,
\\
L_m
&=&\sum_{I=c,d,b,r} a^3 \biggl[ (\dot{v}_I-3H c_I^2 v_I-\alpha) \delta \rho_I
-\frac{c_I^2}{2(\rho_I+P_I)}\delta \rho_I^2-\frac{\rho_I+P_I}{2a^2} 
\partial_i v_I \left( \partial_i v_I+2\partial_i \chi \right)
-\frac{\rho_I}{2a^2}(\partial_i \chi)^2+\frac{\rho_I}{2} \alpha^2 \nonumber \\
&&\qquad \quad~ 
+\frac{P_I}{2} \left( \zeta+\partial^2 E \right) 
\left( 3\zeta-\partial^2 E \right)+
\left\{ (\rho_I+P_I) \left( \dot{v}_I-3H c_I^2 v_I \right)-\rho_I \alpha 
\right\}\left( 3\zeta+\partial^2 E \right)  \biggr]\,,\\
L_{\rm int} &=&\frac{f}{2}a \left[ (\partial_i \chi)^2+a^2 \left\{ 
2(3\zeta+\partial^2 E)\alpha -\alpha^2+(\zeta+\partial^2 E)(3\zeta-\partial^2 E) 
\right\} \right]+\frac{f_{,Z}}{2}a\left( \partial_i v_d
-\partial_i v_c \right)^2\nonumber \\
&&-a^3 f_{,Z} \left[ \dot{v}_d-\dot{v}_c+3H (v_d-v_c) \right] 
\left( \frac{\delta \rho_d}{\rho_d+P_d}-\frac{\delta \rho_c}{\rho_c+P_c} 
\right)\,.
\ea
Varying the action (\ref{S2}) with respect to the non-dynamical perturbations 
$\alpha$, $\chi$, $v_I$, and $E$ and using the background 
Eqs.~(\ref{back1})-(\ref{back2}), we obtain
\ba
& & 
6H M_{\rm pl}^2 \left( H \alpha-\dot{\zeta} \right)
+\frac{2M_{\rm pl}^2}{a^2} \left( \partial^2 \zeta+H\partial^2 \chi
-a^2 H \partial^2 \dot{E} \right)+\sum_{I=c,d,b,r} \delta \rho_I=0\,,
\label{pereq1}\\
& &
2M_{\rm pl}^2  \left( H \alpha-\dot{\zeta} \right)
-\sum_{I=c,d,b,r} \left( \rho_I+P_I \right) v_I=0\,,\\
& &\dot{\delta \rho}_I+3H \left( 1+c_I^2 \right) \delta \rho_I
+3 \left (\rho_I+P_I \right) \dot{\zeta}
-\frac{1}{a^2} \left( \rho_I+P_I \right) 
\left( \partial^2 v_I+\partial^2 \chi-a^2\partial^2\dot{E} \right)=0\,, 
\quad {\rm for} \quad I=c,d,b,r, \label{delrho}\\
& &
\ddot{\zeta}+3H \dot{\zeta}-H \dot{\alpha}-\left( 3H^2+\dot{H} \right) \alpha
-\frac{1}{2M_{\rm pl}^2} \sum_{I=c,d,b,r}
\left( \rho_I+P_I \right) \left( 3H c_I^2 v_I-\dot{v}_I \right)=0\,.
\label{eqze}
\ea
Variations of the quadratic action .~(\ref{S2}) with respect to $v_c$ and $v_d$ actually 
lead to the coupled differential equations of $\delta \rho_c$ 
and $\delta \rho_d$ containing a dependence on  $f_{,Z}$, 
but solving them for $\dot{\delta \rho_c}$ and $\dot{\delta \rho_d}$ 
gives rise to Eqs.~(\ref{delrho}) with $I=c,d$.
Note that these differential equations for $\delta \rho_c$ and $\delta \rho_d$ also follow from perturbing the continuity Eq.~(\ref{umu}).

Variations of the action (\ref{S2}) with 
respect to the dynamical perturbations 
$\delta \rho_I$ give 
\ba
& &
\dot{v}_c-3H c_c^2 v_c-\alpha-c_c^2 \frac{\delta \rho_c}{\rho_c+P_c}
+\frac{f_{,Z}}{\rho_c+P_c} \left[ \dot{v}_d-\dot{v}_c 
+3H \left( v_d-v_c \right) \right]=0\,,\label{vceq}\\
& &
\dot{v}_d-3H c_d^2 v_d-\alpha-c_d^2 \frac{\delta \rho_d}{\rho_d+P_d}
-\frac{f_{,Z}}{\rho_d+P_d} \left[ \dot{v}_d-\dot{v}_c 
+3H \left( v_d-v_c \right) \right]=0\,,\label{vdeq}\\
& &
\dot{v}_I-3H c_I^2 v_I-\alpha-c_I^2 \frac{\delta \rho_I}{\rho_I+P_I}=0
\qquad({\rm for}~I=b,r)\,.
\ea
The effect of momentum exchange between CDM and DE 
appears as the $f_{,Z}$-dependent terms in Eqs.~(\ref{vceq}) and (\ref{vdeq}). 
We need to combine Eqs.~(\ref{vceq}) and (\ref{vdeq}) 
to solve the differential 
equations for $v_c$ and $v_b$.
Varying Eq.~(\ref{S2}) with respect to $\zeta$ and combining 
it with Eq.~(\ref{eqze}), it follows that 
\be
\alpha+\zeta+\dot{\chi}+H \chi
-a^2 \left( \ddot{E}+3H \dot{E} \right)=0\,.
\label{pereq8}
\ee
As advertised above, the interaction between CDM and DE only 
affects the Euler equations describing the momentum conservation 
of the system. This type of coupling was dubbed pure momentum exchange 
in Ref.~\cite{Pourtsidou:2013nha}. 
It also presents some resemblance with the scenario discussed 
in Ref.~\cite{Simpson:2010vh}, where a possible elastic scattering 
of DE is analysed, and in Ref.~\cite{Asghari:2019qld}, 
where an interaction between CDM and DE proportional to their 
relative velocities is explored. In these two scenarios, 
the effect is governed by the relative velocities of the fluids, 
similar to what happens with a Thomson-like scattering. 
In our scenario, it seems the specific interaction driven
by the relative velocity is not realizable (at least in its simplest formulation), 
but a term proportional to the relative 
acceleration ($\dot{v}_d-\dot{v}_c$) also arises.

\subsection{Gauge-invariant perturbation equations}

The perturbation Eqs.~(\ref{pereq1})-(\ref{pereq8}) can be 
expressed in terms of variables invariant 
under the infinitesimal coodinate transformation 
$t \to t+\xi^{0}$ and $x^{i} \to x^{i}+\delta^{ij} \partial_{j} \xi$. 
We introduce the following gauge-invariant 
combinations \cite{Bardeen:1980kt}
\ba
& &
\Psi=\alpha+\frac{{\rm d}}{{\rm d}t} 
\left( \chi - a^2 \dot{E} \right)\,,\qquad 
\Phi=-\zeta-H \left( \chi - a^2 \dot{E} \right)\,,
\nonumber \\
& &
\delta\rho_{I\rm N}=\delta \rho_I+\dot{\rho}_I \left(\chi-a^2 \dot{E}\right)\,,
\qquad 
v_{I{\rm N}}=v_I+\chi-a^2 \dot{E}\,,
\label{delphiN}
\ea
and rewrite the perturbation equations 
by using these variables\footnote{ 
Compared to the notation used in Refs.~\cite{Kase:2019veo,Kase:2020hst}, 
the sign of $\Phi$ is opposite.}. 
In the following, we will switch to the Fourier space with 
a comoving wavenumber $k$. 
Then, all the gauge-dependent quantities like $\chi$ and $E$ 
disappear from Eqs.~(\ref{pereq1})-(\ref{pereq8}) and 
we end up with the following equations
\ba
\hspace{-0.7cm}
& &
6HM_{\rm pl}^2 \left( \dot{\Phi}+H \Psi \right)
+\frac{2k^2}{a^2} M_{\rm pl}^2 \Phi
+\sum_{I=c,d,b,r}\delta \rho_{I{\rm N}}=0\,,\label{peq1}\\
\hspace{-0.7cm}
& &
2M_{\rm pl}^2 \left( \dot{\Phi}+H \Psi \right)
-\sum_{I=c,d,b,r} \left( \rho_I+P_I \right) v_{I{\rm N}}=0\,,
\label{peq2}\\
\hspace{-0.7cm}
& &
\dot{\delta \rho}_{I{\rm N}}+3H \left( 1+c_I^2 \right) \delta \rho_{I{\rm N}}
-3(\rho_I+P_I) \dot{\Phi}+\frac{k^2}{a^2} \left( \rho_I+P_I 
\right) v_{I{\rm N}}=0\,,\quad {\rm for} \quad I=c,d,b,r, 
\label{tdelrho}\\
\hspace{-0.7cm}
& &
\ddot{\Phi}+3H \dot{\Phi}+H \dot{\Psi}+\left( 3H^2+\dot{H} \right) \Psi
+\frac{1}{2M_{\rm pl}^2} \sum_{I=c,d,b,r}\left( \rho_I+P_I \right)
\left( 3H c_I^2 v_{I{\rm N}}-\dot{v}_{I{\rm N}} \right)=0\,,\\
\hspace{-0.7cm}
& &
\dot{v}_{c{\rm N}}-3H c_c^2 v_{c{\rm N}}-\Psi
-\frac{(\rho_d+P_d)[c_c^2 \delta \rho_{c{\rm N}}+
3H f_{,Z}\{(1+c_c^2)v_{c{\rm N}}-(1+c_d^2)v_{d{\rm N}}\}]
-f_{,Z}(c_c^2 \delta \rho_{c{\rm N}}+c_d^2 \delta \rho_{d{\rm N}})}
{(\rho_c+P_c)(\rho_d+P_d)-f_{,Z}(\rho_c+P_c+\rho_d+P_d)}=0\,,
\label{tvcN}\\
\hspace{-0.7cm}
& &
\dot{v}_{d{\rm N}}-3H c_d^2 v_{d{\rm N}}-\Psi
-\frac{(\rho_c+P_c)[c_d^2 \delta \rho_{d{\rm N}}+
3H f_{,Z}\{(1+c_d^2)v_{d{\rm N}}-(1+c_c^2)v_{c{\rm N}}\}] 
-f_{,Z}(c_c^2 \delta \rho_{c{\rm N}}+c_d^2 \delta \rho_{d{\rm N}})}
{(\rho_c+P_c)(\rho_d+P_d)-f_{,Z}(\rho_c+P_c+\rho_d+P_d)}=0\,,
\label{tvdN}\\
\hspace{-0.7cm}
& &
\dot{v}_{I{\rm N}}-3H c_I^2 v_{I{\rm N}}-\Psi
-\frac{c_I^2}{\rho_I+P_I}\delta \rho_{I{\rm N}}=0\,,\quad {\rm for} \quad I=b,r, \\
\hspace{-0.7cm}
& & \Psi=\Phi\,.
\ea
For the derivation of Eqs.~(\ref{tvcN}) and (\ref{tvdN}), we explicitly solved Eqs.~(\ref{vceq}) and 
(\ref{vdeq}) for $\dot{v}_{c{\rm N}}$ and $\dot{v}_{d{\rm N}}$, respectively. 
\subsection{Stability conditions}

Let us derive conditions for the absence of ghost and Laplacian instabilities 
in the small-scale limit (which is still in the regime where 
the linear perturbation theory is valid). 
Since these conditions are independent of the 
gauge choices, we choose the flat gauge characterized by 
\be
\zeta=0\,,\qquad E=0\,.
\ee
Then, the gauge-invariant density perturbation 
$\delta \rho_{I{\rm f}}=\delta \rho_I-\dot{\rho}_I \zeta/H$ and 
velocity potential $v_{I{\rm f}}=v_I-\zeta/H$ are equivalent to 
$\delta \rho_I$ and $v_{I}$, respectively. 
We solve Eqs.~(\ref{pereq1})-(\ref{delrho}) for $\alpha$, $\chi$, and 
$v_I$ ($I=c,d,b,r$) and eliminate these 
non-dynamical perturbations 
from the second-order action (\ref{S2}).
In Fourier space, the second-order action reduces to 
\be
{\cal S}^{(2)}=\int {\rm d}t\,{\rm d}^3k\,a^{3}\left( 
\dot{\vec{\mathcal{X}}}^{t}{\bm K}\dot{\vec{\mathcal{X}}}
-\frac{k^2}{a^2}\vec{\mathcal{X}}^{t}{\bm G}\vec{\mathcal{X}}
-\vec{\mathcal{X}}^{t}{\bm M}\vec{\mathcal{X}}
-\frac{k}{a}\vec{\mathcal{X}}^{t}{\bm B}\dot{\vec{\mathcal{X}}}
\right)\,,
\label{Ss2}
\ee
where ${\bm K}$, ${\bm G}$, ${\bm M}$, ${\bm B}$ 
are $4 \times 4$ matrices, and
\be
\vec{\mathcal{X}}^{t}=\left( 
\delta \rho_{c{\rm u}}/k, 
\delta \rho_{d{\rm u}}/k, 
\delta \rho_{b{\rm u}}/k, 
\delta \rho_{r{\rm u}}/k \right) \,.
\label{calX}
\ee
In the limit that $k \to \infty$, the dominant contributions to 
${\bm K}$ and ${\bm G}$ are given, respectively, by 
\ba
& &
K_{11}=\frac{a^2 (\rho_c+P_c-f_{,Z})}{2(\rho_c+P_c)^2}\,,\qquad 
K_{22}=\frac{a^2 (\rho_d+P_d-f_{,Z})}{2(\rho_d+P_d)^2}\,,\qquad 
K_{12}=K_{21}=\frac{a^2 f_{,Z}}{2(\rho_c+P_c)(\rho_d+P_d)}\,,\nonumber \\
& &
K_{33}=\frac{a^2}{2(\rho_b+P_b)}\,,\qquad 
K_{44}=\frac{a^2}{2(\rho_r+P_r)}\,,
\label{Kcomp}
\\
& &
G_{11}=\frac{a^2 c_c^2}{2(\rho_c+P_c)}\,,\qquad 
G_{22}=\frac{a^2 c_d^2}{2(\rho_d+P_d)}\,,\qquad 
G_{33}=\frac{a^2 c_b^2}{2(\rho_b+P_b)}\,,\qquad 
G_{44}=\frac{a^2 c_r^2}{2(\rho_r+P_r)}\,.
\label{Gcomp}
\ea
The leading-order contributions to the matrix components of 
${\bm M}$ and ${\bm B}$ are of the orders of $k^0$ and $1/k$, 
respectively, so they do not affect the dispersion 
relation in the small-scale limit.

By virtue of the Sylvester criterion, we deduce that ghosts 
are absent under the four conditions $K_{11}>0$, 
$K_{11}K_{22}-K_{12}^2>0$, $K_{33}>0$, 
and $K_{44}>0$, 
which translate to 
\ba 
& &
\rho_c+P_c-f_{,Z}>0\,,\qquad 
\left( \rho_c+P_c \right) \left( \rho_d+P_d \right)
-f_{,Z} \left( \rho_c+P_c+\rho_d+P_d \right)>0\,,\nonumber \\
& &
\rho_b+P_b>0\,,\qquad \rho_r+P_r>0\,.
\label{noghost}
\ea
The latter two are simply the standard weak energy 
conditions of baryons and radiation, but the presence of momentum 
transfer affects the no-ghost conditions of CDM and DE. 
As long as $f_{,Z}<0$ with $\rho_c+P_c>0$ and $\rho_d+P_d>0$, 
there are no ghosts in CDM and DE sectors.

The propagation speed squared $c_s^2$ for the 
high-frequency modes is obtained by solving 
\be
{\rm det} \left| c_s^2 {\bm K}-{\bm G} \right|=0\,.
\ee
Since the baryons and radiation components are decoupled, the matrices are block-diagonal and we have the two following obvious solutions for the above dispersion equation 
$c_b^2=G_{33}/K_{33}$ and $c_r^2=G_{44}/K_{44}$, that coincide with the baryons and radiation propagation speeds. 
The other two solutions corresponding to the coupled $2\times2$ dark sector are given by 
\be
c_s^2=\frac{K_{11}G_{22}+K_{22}G_{11} \pm 
\sqrt{(K_{11}G_{22}+K_{22}G_{11})^2
-4(K_{11}K_{22}-K_{12}^2)G_{11}G_{22}}}
{2(K_{11}K_{22}-K_{12}^2)}\,.
\label{cs2}
\ee
For CDM we have that $c_c^2$ is strongly suppressed,
so we can consider
\be
c_c^2=0\,.
\ee
Since $G_{11}=0$ in this case, the two solutions 
of Eq.~(\ref{cs2}) reduce to
\ba
& &
c_{s1}^2=0\,,\\
& &
c_{s2}^2=c_d^2 \frac{(\rho_c+P_c-f_{,Z})(\rho_d+P_d)}
{\left( \rho_c+P_c \right) \left( \rho_d+P_d \right)
-f_{,Z} \left( \rho_c+P_c+\rho_d+P_d \right)}\,.
\label{eq:soundspeeds}
\ea
We then obtain the sound speed squared $c_{s1}^2$ that vanishes, 
describing the pure CDM modes and the modes with the sound 
speed squared $c_{s2}^2$ that originate from the pressure 
of the DE component and incorporate the extra load sourced 
by the CDM dragging.
The absence of Laplacian instabilities for the coupled system 
requires that
\be
c_{s2}^2 \geq 0\,.
\ee
Incorporating the no-ghost conditions (\ref{noghost}), we find 
the stability condition $c_d^2 (\rho_d+P_d) \geq 0$, 
which holds for $c_d^2 \geq 0$ and $\rho_d+P_d>0$. 
They follow naturally from imposing the null energy condition 
on the DE sector together with the absence of Laplacian instabilities 
in the uncoupled regime.

In summary, under the conditions $\rho_I+P_I>0$ and $c_I^2 \geq 0$ for 
each fluid, there are neither ghost nor Laplacian instabilities 
for $f_{,Z}<0$. 
In this case, we also have the inequality $0 \le c_{s2}^2<c_d^2$.

\section{Particular example: Early dark radiation}
\label{modelsec}

In this section, we will apply the developed general formalism 
to a particular model for the DE sector. 
Before doing so, it is worth noticing that we can actually be 
very general concerning the interaction $f(Z)$. 
As a matter of fact, if we impose that the interaction does not affect the background, we can fix it to be any function such that\footnote{Giving up on this condition would amount to adding a contribution to the cosmological constant, so our Ansatz does not introduce any restriction. It would then suffice to work with the hatted variables introduced 
in Eq.~(\ref{hatrhoP}).} 
\be
f \left( Z=-1 \right)=0\,.
\ee
Furthermore, since the perturbations depend on $f_{,Z}$ alone evaluated on the background, the interaction only introduces an additional constant parameter that we denote 
\be
b\equiv \big(f_{,Z}\big)_{\vert_{Z=-1}}\,.
\ee
Let us notice that the stability conditions are guaranteed to be satisfy if $b<0$. We will encounter this condition again below to be sufficient to avoid Laplacian instabilities.
The CDM is assumed to have an energy density 
of the form,
\be
\rho_c(n_c)=m_c n_c\,,
\ee
where $m_c$ is a constant. 
{}From Eqs.~(\ref{Pdef}) and (\ref{cI}), we have
\be
w_c=0\,,\qquad c_c^2=0\,.
\ee
For the DE fluid, we will consider the following form: 
\be
\rho_d (n_d)=\rho_\Lambda\left(1+r_0 n_d^{1+\cs^2}\right)\,,
\ee
where $\rho_\Lambda$, $r_0$ and $\cs^2$ are positive 
constants whose physical significance will become clear soon.
In this case, we have that the equation of state and 
sound speed squared are given, respectively, by
\be
w_d=-\frac{1 -\cs^2 r_0 n_d^{1+\cs^2}}
{1+r_0 n_d^{1+\cs^2}}\,,\qquad
c_d^2=\cs^2\,,
\ee
together with the pressure
\be
P_d=\rho_{d,n_d}n_d-\rho_d= -\rho_\Lambda\left(1-
\cs^2 r_0 n_d^{1+\cs^2}\right). 
\ee
It is now apparent that the DE can be interpreted as a combination of the cosmological constant given by $\rho_\Lambda$ and the energy density of 
a perfect fluid proportional to $n_d^{1+\cs^2}$.
The positive constant $\cs^2$ corresponds to the propagation speed 
squared of the whole DE sector at all times (as expected since the cosmological constant contribution does not exhibit perturbations). 
{}From Eq.~(\ref{ncon}) the number density 
decreases as $n_d \propto a^{-3}$, so the past asymptotic value 
of $w_d$ is equivalent to $\cs^2$.
After $r_0 n_d^{1+\cs^2}$ drops below $2/(1+3\cs^2)$, 
$w_d$ gets smaller than $-1/3$ and finally approaches 
$-1$. Hence the Universe enters the stage of 
cosmic acceleration at late times.
This means that DE is composed of a perfect fluid with equation of state $w_d=\cs^2$ at early times and a cosmological constant with energy density $\rho_\Lambda$. The parameter $r_0$ then measures the relative fraction of the two components making up the dark sector or, in other words, it fixes the time at which the cosmological constant takes over the dominance in the DE component.

We will further specify the model under consideration by fixing the parameter $\cs^2$. In principle, to avoid conflicts with Early universe constraints such as big bang nucleosynthesis, imposing any 
$\frac13\geq\cs^2\geq 0$ would do the job. 
However, we would like to impose a scaling behaviour 
in the early Universe 
so that we will fix 
\be
c_d^2=\cs^2=\frac{1}{3}\,.
\ee
The reason for this choice is to avoid additional fine-tunings related to the initial conditions. Under these assumptions, the model is completely specified by just two constant parameters, i.e., the coupling $b$ and the fraction of dark radiation in the early Universe that is related to $r_0$. 
We can then write the energy density, pressure, and equation of 
state of the DE fluid, respectively, as 
\be
\rho_d =\rho_\Lambda\left(1+r a^{-4}\right)\,,\qquad 
P_d=-\rho_\Lambda\left(1-\frac{r}{3} a^{-4}\right)\,,\qquad
w_d=-\frac{1-ra^{-4}/3}{1+ra^{-4}}\,,
\label{rhoPwd}
\ee
where we defined $r \equiv r_0 n_{d0}^{4/3}$ 
with $n_{d0}$ today's DE number density.
We introduce today's density parameter of each 
matter species 
as $\Omega_I=\rho_I/(3M_{\rm pl}^2 H_0^2)$.
The density parameters of dark radiation and 
cosmological constant have the relation
$\Omega_{dr}=r\Omega_\Lambda$. 
The initial fraction of dark radiation as compared to the standard radiation is given by  $\Omega_{dr}/\Omega_r=(\Omega_\Lambda/\Omega_r)r$. Since $\Omega_\Lambda/\Omega_r \simeq 10^4$, in order to avoid having a dominant dark radiation component in the early Universe, we need to require 
that $r\lesssim 10^{-4}$. Furthermore, BBN constraints only allow for a variation of $\sim 10\,\%$ in the Hubble expansion rate which translates into $(\Delta H/H)_{\rm BBN}\simeq (\Delta \rho/\rho)_{\rm BBN}/2 
\lesssim 10^{-1}$. 
This imposes an upper bound in the fraction of dark radiation that gives the limit $r\lesssim 10^{-5}$. This is a rough estimate since the abundances of primordial elements exhibit an exponential dependence on $H$ through the Boltzmann factor, so a more conservative limit that safely evades the BBN constraints can be taken as $r\lesssim 10^{-6}$.  

Although the dark radiation component is completely negligible at late times and the cosmological constant gives the dominant contribution to $\rho_d$, 
the small fraction of dark radiation in the pre-recombination era goes in the correct direction to ease the $H_0$ tension since it reduces the sound horizon at recombination and, therefore, we need to increase $H_0$ to keep the position of CMB acoustic peaks. Usually, this effect introduces further conflict with the $\sigma_8$ tension because a higher value of $H_0$ gives rise to a larger value of $\sigma_8$ \cite{Hill:2020osr}. In the present scenario, however, we will see that the interaction of CDM with the DE sector prevents the growth of structures so the additional radiation does not worsen the $\sigma_8$ tension, but it can actually resolve it. This is in line with the results obtained from models incorporating a dark radiation component with a non-negligible cross section with CDM particles (see e.g. \cite{Buen-Abad:2015ova,Lesgourgues:2015wza,Buen-Abad:2017gxg,Raveri:2017jto,Blinov:2020hmc}).

To study the evolution of perturbations, we will introduce the following gauge-invariant quantities describing the density contrasts and velocity potentials of the corresponding components:
\be
\delta_{I{\rm N}} \equiv \frac{\delta \rho_{I{\rm N}}}{\rho_I}\,,\qquad
\theta_{I{\rm N}} \equiv \frac{k^2}{a}v_{I{\rm N}}\,.
\ee
{}From Eqs.~(\ref{tdelrho}), (\ref{tvcN}), and (\ref{tvdN}), the 
perturbations $\delta_{c{\rm N}}$, $\delta_{d{\rm N}}$, $\theta_{c{\rm N}}$, and $\theta_{d{\rm N}}$ obey the following differential equations 
\ba
& &
\delta_{c{\rm N}}'=3\Phi'-\theta_{c{\rm N}}\,,
\label{delceq}\\
& &
\delta_{d{\rm N}}'=-3{\cal H} \left( c_d^2-w_d \right) 
\delta_{d{\rm N}}+3(1+w_d)\Phi'-(1+w_d)\theta_{d{\rm N}}\,,
\label{deldeq}\\
& &
\theta_{c{\rm N}}'=-{\cal H}\theta_{c{\rm N}}
+k^2 \Phi 
+b\frac{3 {\cal H} (1+w_d) \rho_d 
[\theta_{c{\rm N}}-(1+c_d^2) \theta_{d{\rm N}}]
-k^2  c_d^2 \rho_d \delta_{d{\rm N}}}
{(1+w_d)\rho_d(\rho_c-b)-b\rho_c}\,,
\label{thetaceq} \\
& &
\theta_{d{\rm N}}'={\cal H} (3c_d^2-1) \theta_{d{\rm N}}+k^2 \Phi 
+\frac{\rho_c[k^2 c_d^2 \rho_d \delta_{d{\rm N}}+
3 {\cal H} b\{(1+c_d^2) \theta_{d{\rm N}}-\theta_{c{\rm N}}\}]
-k^2 b c_d^2 \rho_d \delta_{d{\rm N}}}
 {(1+w_d)\rho_d(\rho_c-b)-b\rho_c}\,,
 \label{thetadeq}
\ea
where a prime represents the derivative with 
respect to the conformal time $\tau=\int a^{-1} {\rm d}t$, 
and ${\cal H}=aH$.

Before moving on to analysing the evolution of the system governed by these equations, let us notice that we could have been more general and allowed for a dependence on the number densities in the couplings. If we consider a more general coupling function of the form $\tilde{f}(n_c,n_d,Z)={\cal F}(n_c,n_d) f(Z)$, then the perturbation equations would read exactly the same but now with a time-dependent coupling obtained via the replacement $b\rightarrow {\cal F}(n_c,n_d)b$ in the above equations. However, if we assume an analytical function ${\cal F}$ so that 
${\cal F}=\sum_{i,j\geq0} {\cal F}_{ij} n_c^in_d^j$, at late times only the component 
${\cal F}_{00}={\cal F}(0,0)$ is relevant. By absorbing its value into the value of $b$ we would be back to our equations. In this work we are interested in having effects at late times where DE is relevant, so we will stick to our simple scenario with constant $b$, but extensions to other models with effect at earlier times can be straightforwardly studied within our framework. 

\section{Linear growth of structures}\label{Sec:growth}

Equipped with the perturbation equations of motion, 
we can proceed to study how the CDM clustering occurs 
in the presence of momentum exchange with the DE fluid. 
As it should be obvious, since the new terms in the equations are determined by the relative velocities of CDM and DE (or its derivatives), there are no effects whenever the perturbations evolve in an adiabatic regime. This occurs for the super-Hubble modes and for the adiabatic initial conditions generated during inflation, so all the differences are expected to take place in the sub-Hubble regime. In this respect, this evolution is guaranteed to occur by the existence of the conserved Weinberg adiabatic mode as the dominant solution. It should be checked that the interaction terms do not introduce additional modes that grow with respect to the conserved one, but this is trivial since, as commented above, adiabatic modes do not contribute to the interaction. Thus, any deviation from the standard super-Hubble evolution driven by the interaction terms must be caused by 
non-adiabatic modes.

\subsection{Velocity potentials}
\label{velosec}

Let us first look at the evolution of velocity potentials 
in the regime where $|b|\gg\rho_c$ to study the effect of the interaction. 
In this case, there are two possibilities, either $\rho_d\gg\rho_c$ or $\rho_d\ll\rho_c$. The former only happens marginally at very late times when 
DE dominates, while at early times it could happen if the condition 
$|b| \gg\rho_c$ is satisfied before radiation-matter equality. 
The latter corresponds to the core of matter domination, 
so let us look at this case first. Under the conditions 
$|b| \gg\rho_c$ and $\rho_d\ll\rho_c$, the coupled 
Euler Eqs.~(\ref{thetaceq}) and (\ref{thetadeq}) 
can be written as
\begin{eqnarray}
\frac{\dd }{\dd N}\begin{bmatrix}
\theta_{c}  \\
\theta_{d} 
\end{bmatrix}\simeq
\begin{bmatrix}
-1&3(1+w_d)(1+c_d^2)R_d\\
3&-4
\end{bmatrix}
\begin{bmatrix}
\theta_{c}  \\
\theta_{d} 
\end{bmatrix}
+\frac{1}{\mathcal{H}}\begin{bmatrix}
1  \\
1
\end{bmatrix} S_k\,,
\label{eq:Eulerlargeb}
\end{eqnarray}
where $N=\int H {\rm d}t$ is the e-folding 
number, and we have introduced the quantities
\be
R_d=\frac{\rho_d}{\rho_c}\,,\qquad 
S_k=R_dc_d^2k^2 \delta_{d}+k^2 \Phi\,.
\ee
These equations are also valid in the synchronous gauge by simply setting $\Phi=0$ in the source $S_k$, so we have dropped the 
subscript ``N'' referring to the Newtonian gauge. A remarkable property of the regime where the above equations are valid is the explicit disappearance of the coupling parameter $b$. Of course, this does not mean that the interaction does not play any role. Firstly, the evolution of the perturbations is modified by the interaction, although in a manner that is insensitive to the value of $b$. Secondly, the time at which we enter the regime with $|b|\gg\rho_c$ does depend on the explicit value of $b$, 
so the amount of time during which the perturbations 
are subject to the modified evolution is sensitive to $b$ 
and this can impact the matter power spectrum 
in a $b$-dependent manner as we will show below. 

From Eq.~\eqref{eq:Eulerlargeb}, we can obtain a universal and remarkably simple relation between the two velocities. Let us first notice that the eigenvalues of the homogeneous system (with $S_k=0$) are $\lambda_1=-1$ and $\lambda_2=-4$ up to corrections of order $(1+w_d)R_d$, while the eigenvectors are $\vec{v}_1=(1,1)$ and $\vec{v}_2=(0,1)$, again up to corrections of order $(1+w_d)R_d$. Since the source term is proportional to the second eigenvector, the solution, in the limit $R_d\ll1$, is given by
\be
\vec{\theta}\simeq \left( C_{1k}+\int^\tau aS_k \dd \tilde{\tau} \right) \vec{v}_1 a^{-1}
+C_{2k}\vec{v}_2 a^{-4}\,,
\label{eq:thetaslargeb}
\ee
with $\vec{\theta}=(\theta_c,\theta_d)$ and $C_{1k}$, $C_{2k}$ the integration constants. So far we have not taken any sub-Hubble limit, so imposing adiabatic initial conditions for the modes that enter the considered regime being super-Hubble will have $C_{2k}=0$. Notice that the whole $k$-dependence 
in the solution (\ref{eq:thetaslargeb}) comes from the source term $S_k$. In any case, since the mode 
$C_{2k}a^{-4}$ decreases faster than the mode 
$C_{1k}a^{-1}$, the former will be negligible at late times. Then, the solution \eqref{eq:thetaslargeb} shows that the peculiar velocities are actually equal 
to each other (up to corrections of order $R_d$ 
that we have neglected), i.e., 
\be
\theta_c\simeq\theta_d\,,
\label{thetacd}
\ee
in this regime. Furthermore, this property does not depend on the specific evolution of the perturbations and it holds in any gauge. This should not come as a surprise, since in the strongly interacting regime with $|b| \gg \rho_c$ the two fluids are expected to move together. This is analogous to the CMB photons tightly coupled to baryons due to Thomson scattering before recombination. 

\subsection{Density contrasts}
\label{densec}

Let us now turn our attention to the process of (linear) structure formation and how the interaction affects the evolution of CDM and DE density contrasts. Since this process mostly takes place during matter domination, we again assume that the Universe is in the matter-dominated regime. As for the interaction, we make an assumption that 
$|b|\gg (1+w_d) \rho_d$ but not necessarily $|b|\gg \rho_c$. Under this assumption we can approximate the denominators in the interaction term of the perturbation Eqs.~(\ref{thetaceq}) and (\ref{thetadeq}), by $(1+w_d)\rho_d(\rho_c-b)-b\rho_c\simeq-b\rho_c$ regardless the hierarchy between $\rho_c$ and $b$.  

By using Eqs.~(\ref{peq1}) and (\ref{peq2}), the gravitational potential $\Phi$ during the matter dominance can be expressed as
\be
k^2\Phi=-\frac32 \cH^2\left[\delta_{c{\rm N}}
+3\cH\frac{\theta_{c{\rm N}}}{k^2}
+\left( \delta_{d{\rm N}}+3(1+w_d)\cH\frac{\theta_{d{\rm N}}}{k^2}
\right) R_d\right]\,,
\ee
where we ignored the contribution of baryon perturbations.
Since we are interested in sub-Hubble modes, we can neglect the effects of the peculiar velocities on $\Phi$. On the other hand, at early times when the interaction is negligible, the small pressure of CDM favours its clustering as opposed to the DE sector, where the pressure prevents its clustering and keeps it more homogeneous. 
In the regime $R_d\ll1$, we can neglect the DE contribution 
to $\Phi$, so we have the standard relation
\be
k^2\Phi \simeq -\frac32\cH^2\delta_{c{\rm N}}\,.
\ee

As we showed in Eq.~(\ref{thetacd}), the CDM and DE velocities are equal in the strong coupling regime. Prior to this regime, the pressure of the DE component prevents the appearance of large peculiar velocities, whereas the CDM component tends to fall into the gravitational wells in the sub-Hubble regime. Thus, we will also assume that the DE peculiar velocity does not 
exceed the CDM one.  

Under the discussed conditions, i.e., 
$|b|\gg(1+w_d)\rho_d$, $k^2\gg \cH^2$, 
$\theta_{d{\rm N}} \lesssim \theta_{c{\rm N}}$, 
and $R_d\ll1$, the perturbation Eqs.~(\ref{delceq})-(\ref{thetadeq}) 
in the Newtonian gauge can be written as
\ba
& &
\delta_{c{\rm N}}'\simeq-\theta_{c{\rm N}}\,,\\
& &
\delta_{d{\rm N}}'\simeq-3{\cal H} \left( c_d^2-w_d \right) 
\delta_{d{\rm N}}+\frac92(1+w_d)\frac{\cH^3}{k^2}\delta_{c{\rm N}}-(1+w_d)\left(\theta_{d{\rm N}}-\frac{9\cH^2}{2k^2}\theta_{c{\rm N}}\right)\,,\\
& &
\theta_{c{\rm N}}'\simeq-\cH\theta_{c{\rm N}}-\frac32\cH^2\delta_{c{\rm N}}+R_dc_d^2k^2\delta_{d{\rm N}}\,,\\
& &
\theta_{d{\rm N}}'\simeq-4{\cal H}\theta_{d{\rm N}}+3{\cal H}\theta_{c{\rm N}}-\frac32\cH^2\delta_{c{\rm N}}+\left(1-\frac{\rho_c}{b}\right)R_dc_d^2k^2\delta_{d{\rm N}}\,,
\ea
where we have used that $\cH'=-\cH^2/2$ as it corresponds to matter domination. These equations can be combined to obtain a system of two coupled oscillators describing the evolution of the density contrasts. We follow the usual procedure of taking derivatives of the continuity equations and removing the peculiar velocities by using the Euler and continuity equations. By doing so, we obtain
\ba
& &
\delta_{c{\rm N}}''+\cH\delta_{c{\rm N}}'-\frac32\cH^2\delta_{c{\rm N}} \simeq -R_dc_d^2k^2\delta_{d{\rm N}}\,,\label{eq:deltacoscillator}
\\
& &
\delta_{d{\rm N}}''+\Big[4+6(c_d^2-w_d)\Big]
\cH\delta_{d{\rm N}}'+\left[(1+w_d)
\left(1-\frac{\rho_c}{b}\right)R_dc_d^2k^2
+\frac{3}{2}(13+6c_d^2)(c_d^2-w_d)\cH^2\right]\delta_{d{\rm N}} 
\nonumber \\
&&\simeq\frac32(1+w_d)\cH\Big(2\delta_{c{\rm N}}'+\cH\delta_{c{\rm N}}\Big)\,.
\label{eq:deltadoscillator}
\ea
The equation for $\delta_{c{\rm N}}$ decouples 
if $|R_dc_d^2k^2\delta_{d{\rm N}}| \ll |\cH^2\delta_{c{\rm N}}|$. 
In this regime, the CDM density contrast 
grows as usual $\delta_{c{\rm N}}\propto a$ 
(with a subdominant decaying mode). 
The DE density contrast then evolves with 
an effective propagation speed squared given by 
\be
c_{\rm eff}^2=(1+w_d)\left( 1-\frac{\rho_c}{b} 
\right)R_d c_d^2\,,
\label{ceff}
\ee
which coincides with Eq.~\eqref{eq:soundspeeds} 
in the corresponding regime with $|b|\gg(1+w_d)\rho_d$. 
This determines the critical wavenumber corresponding 
to the effective DE sound horizon as
\be
k_s=\frac{{\cal H}}{c_{\rm eff}}\,.
\label{ksdef}
\ee
We again obtain the condition $b<0$ to guarantee 
the absence of Laplacian instabilities. 
Let us estimate the evolution of $c_{\rm eff}$ and $k_s$ 
during the matter dominance (${\cal H} \propto a^{-1/2}$ 
with $a \propto \tau^2$). Since 
$(1+w_d)R_d=(\rho_d+P_d)/\rho_c \propto a^{-1}$, 
we find
\ba
& &
c_{\rm eff} \propto a^{-2}\,,\qquad~~ 
k_s \propto a^{3/2} \propto \tau^3\,,
\qquad {\rm for}~~\rho_c/|b| \gg 1\,,
\label{ceff1}\\
& &
c_{\rm eff} \propto a^{-1/2}\,,\qquad 
k_s={\rm constant}\,,
\qquad~{\rm for}~~\rho_c/|b| \ll 1\,.
\label{ceff2}
\ea
On the de Sitter solution (${\cal H}=aH \propto a$), 
we have the dependence $k_s \propto a^{3/2}$ 
in the regime $\rho_c/|b| \ll 1$.

For scales outside the effective DE sound horizon, i.e.,
\be
k \ll k_s\,,
\ee
it is easy to check that the adiabatic mode 
\be
\delta_{d{\rm N}}^{\rm ad}=(1+w_d)\delta_{c{\rm N}}
\label{adimode}
\ee
is a solution\footnote{At early times when $w_d\simeq c_d^2$ the DE effective mass becomes very small. In that case, our discussion is still valid because we will have $\delta_{d{\rm N}}''\sim \cH^2\delta_{d{\rm N}}$.}. 
Since the effective mass and friction matrices 
in Eq.~(\ref{eq:deltadoscillator}) have non-negative 
real eigenvalues, the solutions of the homogeneous 
equation ($\delta_{d{\rm N}}^{\rm H}{}''
+4\cH\delta_{d{\rm N}}^{\rm H}{}' \simeq 0$) 
do not grow and the above adiabatic mode gives 
the dominant contribution to $\delta_{d{\rm N}}$. 
Furthermore, for this adiabatic solution, 
we can write the aforementioned condition for the decoupling 
of $\delta_{c{\rm N}}$ as
\be
\xi \equiv \frac{R_d c_d^2k^2\delta_{d{\rm N}}^{\rm ad}}{\cH^2\delta_{c{\rm N}}}=\frac{(1+w_d)R_dc_d^2k^2}{{\cH^2}}
=\left( \frac{k}{k_s} \right)^2 
\frac{1}{1-\rho_c/b} \ll 1\,.
\label{ratio}
\ee
As we showed in Eqs.~(\ref{ceff1}) and (\ref{ceff2}), 
the ratio $\xi$ is constant during the matter dominance 
irrespective of the values of $\rho_c/b$. 
Even after the onset of cosmic acceleration, $\xi$ 
decreases due to the increase of ${\cal H}$.
Thus, for $k \ll k_s$, the adiabatic mode (\ref{adimode}) 
is the solution throughout 
the cosmological evolution from the matter dominance to today, 
provided that $\xi \ll 1$ at early times.

For modes inside the effective DE sound horizon, i.e., 
\be
k \gg k_s\,,
\ee
the adiabatic evolution for the DE density contrast ceases. 
In that regime, there is a rapidly oscillating mode of
$\delta_{d{\rm N}}$ induced by the large DE pressure 
associated with the effective DE sound speed squared 
(\ref{ceff}). This corresponds to the homogeneous solution 
$\delta_{d{\rm N}}^{\rm H}$ to Eq.~(\ref{eq:deltadoscillator}).
Since the CDM does not have pressure or, equivalently, its sound speed vanishes, the presence of $\delta_{c{\rm N}}$-dependent 
terms on the right hand-side of Eq.~(\ref{eq:deltadoscillator}) 
gives rise to slow modes $\delta_{d{\rm N}}^{\rm slow}$
which do not exhibit fast oscillations.
For these slow modes, we can neglect the derivatives of $\delta_{d{\rm N}}$ in Eq.~\eqref{eq:deltadoscillator} relative 
to the Laplacian term, such that 
\be
\left(1-\frac{\rho_c}{b}\right)R_d c_d^2k^2 
\delta_{d{\rm N}}^{\rm slow}\simeq \frac32\cH\Big(2\delta_{c{\rm N}}'+\cH\delta_{c{\rm N}}\Big)\,.
\label{eq:subhorizonslow}
\ee
Thus, the general solution to Eq.~\eqref{eq:deltadoscillator} 
can be expressed as
\be
\delta_{d{\rm N}}=
\delta_{d{\rm N}}^{\rm H}+
\delta_{d{\rm N}}^{\rm slow}\,.
\label{deldsum}
\ee
In the regime where the condition 
\be
|\delta_{d{\rm N}}^{\rm slow}| \gg 
|\delta_{d{\rm N}}^{\rm H}|
\ee
is satisfied, we can replace $\delta_{d{\rm N}}$ in Eq.~(\ref{eq:deltacoscillator}) with $\delta_{d{\rm N}}^{\rm slow}$ and exploit the relation (\ref{eq:subhorizonslow}). 
Then, the CDM density contrast obeys 
the following decoupled equation
\be
\delta_{c{\rm N}}''+\cH\left(1+\frac{3}{1-\rho_c/b}\right)\delta_{c{\rm N}}'-\frac32\left(1-\frac{1}{1-\rho_c/b}\right)\cH^2\delta_{c{\rm N}}=0\,.
\label{delcNeq}
\ee
This equation shows that, for $\rho_c \gg |b|$, the CDM density 
contrast deviates from the standard growing-mode solution 
($\delta_{c{\rm N}}\propto a$) only by a small correction 
of order $b/\rho_c$.

In the regime where the interaction between CDM and DE is sufficiently strong ($|b|\gg\rho_c$), 
Eq.~(\ref{delcNeq}) is approximately given by 
\be
\delta_{c{\rm N}}''+4\cH \delta_{c{\rm N}}'
\simeq 0\,,
\ee
%
During the matter dominance ($a \propto \tau^2$), the 
solution to this equation is the sum of a constant mode 
$c_1$ plus a decaying mode $c_2 a^{-7/2}$, i.e., 
\be
\delta_{c{\rm N}}\simeq c_1+c_2 a^{-7/2}\,.
\label{delcfro}
\ee
This result shows how the CDM density contrast freezes 
and it maintains the amplitude with which it entered this regime. Thus, the interaction leads to a late-time suppression 
for the structure formation. 
For this constant mode, we can resort to
Eq.~\eqref{eq:subhorizonslow} to obtain the DE density contrast
\be
\frac{\delta_{d{\rm N}}}{1+w_d} \simeq
\frac{3\cH^2}{2(1+w_d)R_d c_d^2k^2}\delta_{c{\rm N}}
=\frac{3}{2} \left( \frac{k_s}{k} \right)^2 
\delta_{c{\rm N}}\,.
\label{deldfro}
\ee
Since $k_s$ does not vary in time in the strong coupling regime 
of matter era, $\delta_{d{\rm N}}/(1+w_d)$ is constant, 
with the suppression of order $(k_s/k)^2$ in comparison 
to $\delta_{c{\rm N}}$.

It is interesting to notice that the condition for the modes being outside or inside the sound horizon does not depend on $b$. This in turn implies that the suppression of the CDM density contrast does not directly depend on the precise value of $b$. 
It does however depend on $b$ because the strong coupling regime starts at the time $\tau_\star$ determined by the condition $|b|=\rho_c(\tau_\star)$. This gives
\be
a_\star=\left(\frac{|b|}{\rho_{c,{\rm end}}}
\right)^{-1/3} a_{\rm end}\,,
\ee
where $a_{\rm end}$ and $\rho_{c,{\rm end}}$ are 
the scale factor and the CDM density at some final time. 
Technically, this would only give the suppression up to 
the end of matter domination. However, it gives a very good approximation to the suppression today by extrapolating to the DE domination.
Thus, we can straightforwardly compute the suppression of 
the CDM contrast due to the interaction with respect to the non-interacting case by simply scaling the suppression from $\tau_\star$ until $\tau_{\rm end}$ as follows:
\be
\frac{\delta_c}{\delta_c^{b=0}}
=\frac{a_\star}{a_{\rm end}}=
\left(\frac{|b|}{\rho_{c,{\rm end}}}\right)^{-1/3}\,.
\label{rdelta}
\ee
This shows that the suppression has a mild dependence
on the interaction parameter $b$. 
Furthermore, we verify that the ratio (\ref{rdelta}) 
does not depend on $r$, i.e., on the initial fraction of dark radiation, but it only fixes the sound horizon scale that determines which scales undergo a suppressed clustering. 
By using Eq.~(\ref{rhoPwd}), we can easily obtain that the wavenumber (\ref{ksdef}) 
associated with the effective DE sound horizon has the dependence 
\be
k_s \propto \frac{1}{\sqrt{r}}\,.
\label{kses}
\ee
Thus, the CDM density contrast $\delta_c$ 
on scales below this DE sound horizon would exhibit a clustering suppression $\propto |b|^{-1/3}$, while the modes with $k<k_s$ should evolve as in the non-interacting case. 
For the matter power spectrum, we will then have a suppression $\propto |b|^{-2/3}$ for $k> k_s$ and no effects for $k<k_s$. Let us notice that this suppression only affects the CDM component, but not the baryons so in the total matter spectrum the suppression will be slightly milder.

\section{Numerical solutions}

After having obtained analytical solutions for the evolution 
of perturbations in the relevant regimes, we will corroborate 
our findings by numerically solving 
the full system of equations. 
Since we are mainly interested in the evolution of 
perturbations during the matter era 
to show the suppression of CDM clustering 
induced by the momentum transfer, we will focus on 
the post-recombination era well-inside matter domination and will follow the evolution of the perturbations until today where DE dominates. 
The background Friedmann equation can then be well approximated by
\be
\cH^2\simeq\frac{8\pi G}{3}a^2
\left( \rho_c+\rho_b+\rho_\Lambda \right)\,,
\ee
where we have neglected the contribution of dark radiation 
to $\rho_d$. 
For the background, we will fix the value of the DE density 
parameter to be $\Omega_\Lambda=8\pi G \rho_\Lambda/(3H_0^2)=0.7$.
Then, the parameter $r$ in $\rho_d$ fixes the initial fraction of 
dark radiation to that of standard radiation and to have 
$\rho_d\lesssim 10^{-2}\rho_r$ in the early Universe 
we need to require that $r\lesssim 10^{-6}$.

To solve the perturbation equations of motion,  we choose 
the Newtonian gauge and omit the subscript ``N'' 
in the following discussion.
We take the initial conditions of density contrasts 
and velocity potentials as $\delta_{c,{\rm ini}}=1$,  
$\delta_{d,{\rm ini}}=0$, and 
$\theta_{c,{\rm ini}}=0$, $\theta_{d,{\rm ini}}=0$.
The reason for this choice is that we would like to obtain the transfer function for $\delta_c$. Since the perturbations before entering the regime when the interaction becomes effective evolve in the standard manner, 
the transfer function will directly give the effect on the matter power spectrum due to the interaction. In general, the transfer matrix has off-diagonal components that might contribute to the modification in the matter power spectrum. However, these extra contributions are expected to be small, so today's CDM contrast $\delta_{c,0}$ will give the dominant contribution to the total CDM power spectrum.
We will come back to this point later and confirm it numerically. For the moment, it is sufficient to notice that this choice of initial conditions will not affect the subsequent perturbation dynamics since the system rapidly evolves towards the attractor solution driven by the growing mode of $\delta_c$.

\begin{figure}[!t]
\includegraphics[width=0.49\textwidth]{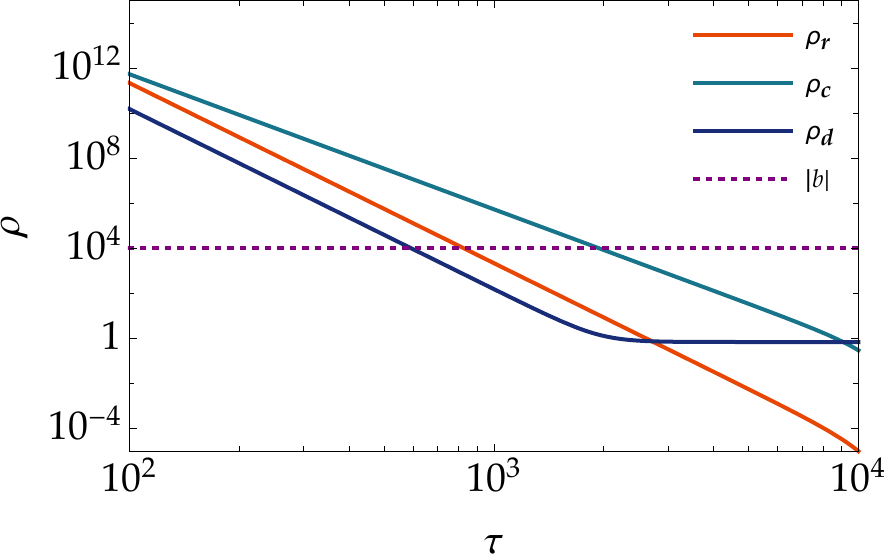}
\includegraphics[width=0.49\textwidth]{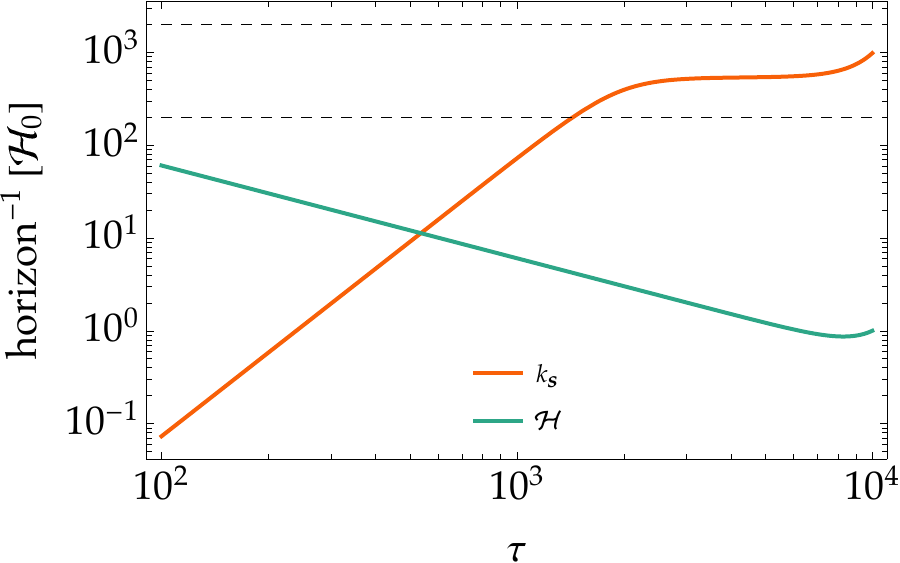}
\vspace{0.3cm}
\includegraphics[width=0.49\textwidth]{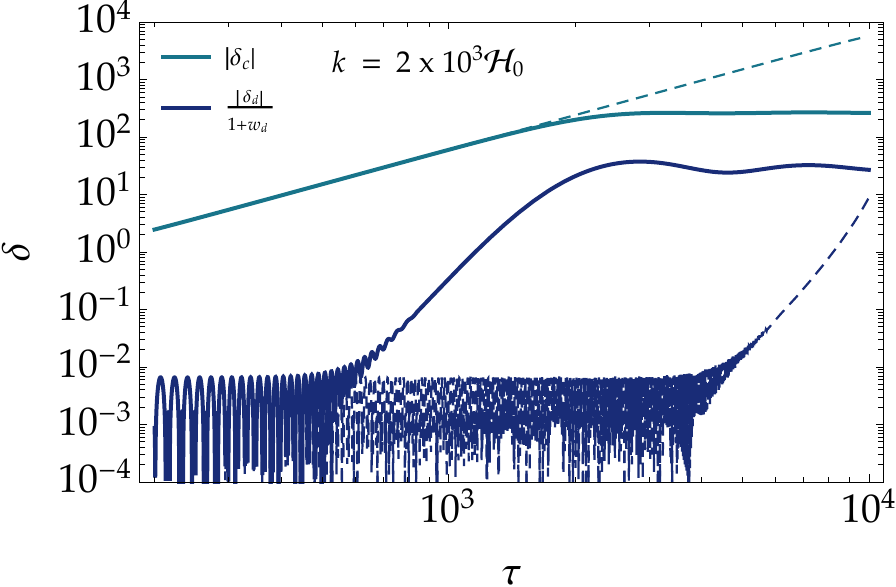}
\includegraphics[width=0.49\textwidth]{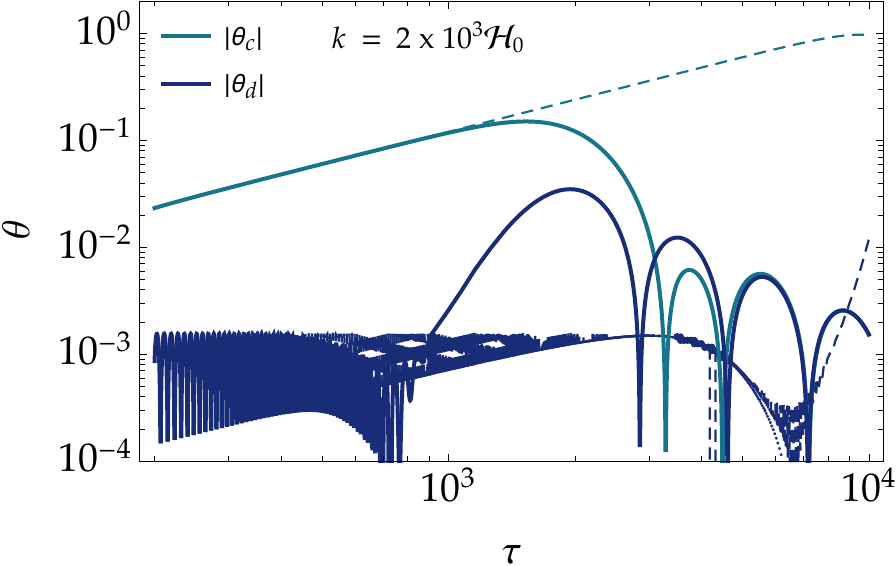}
\vspace{0.3cm}
\includegraphics[width=0.49\textwidth]{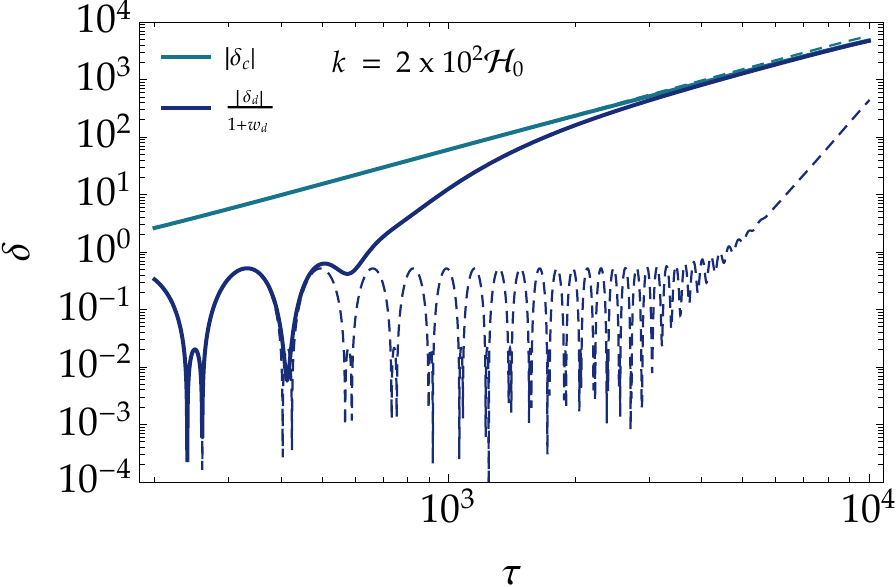}
\includegraphics[width=0.49\textwidth]{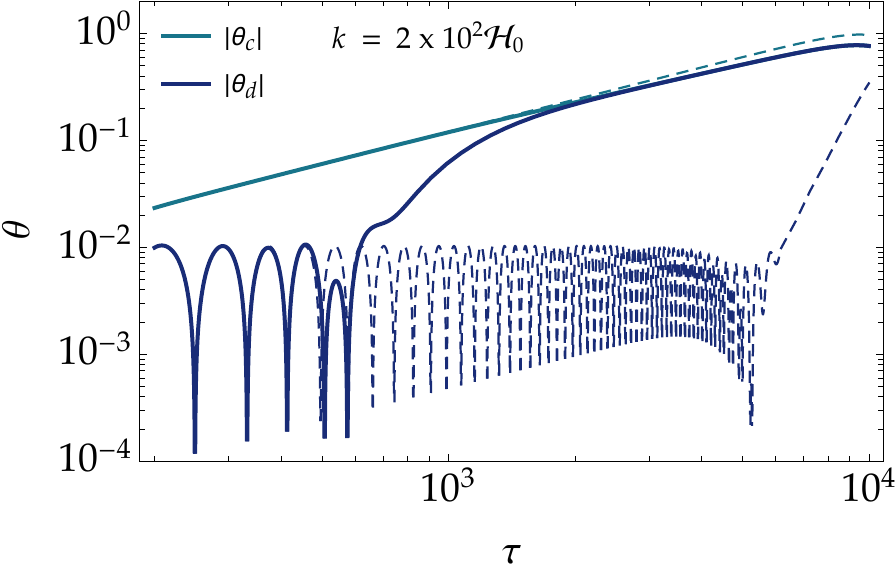}
\caption{In the {\bf upper left panel}, we show the evolution 
of the energy densities $\rho_r$, $\rho_c$, and $\rho_d$ 
as well as the interaction constant 
$b=-10^4\rho_0$ for $r=10^{-6}$. 
The {\bf upper right panel} corresponds to the evolution 
of $k_s$ and $\cH$ normalised by today's Hubble constant $\cH_0$. 
We also show two $k$-modes (horizontal dashed-lines) 
as representatives of a mode that never crossed the DE sound horizon ($k=2\times 10^{3}\cH_0$) and one that was outside the DE sound horizon ($k=2\times 10^{2}\cH_0$) in the strong coupling regime $|b| \gg\rho_c$. 
Since $k_s$ is constant in the regime $|b| \gg\rho_c$ 
during the matter era, there is no horizon crossing of 
the different Fourier modes. 
In the {\bf middle and lower panels}, we plot the evolution 
of the density contrasts and velocity potentials for the modes 
$k=2\times 10^{3}\cH_0$ and $k=2\times 10^{2}\cH_0$, 
respectively, with the initial conditions 
$\delta_{c,{\rm ini}}=1$ and 
$\delta_{d,{\rm ini}}=\theta_{c,{\rm ini}}=\theta_{d,{\rm ini}}=0$.
The evolution for the non-interacting case ($b=0$) 
is also plotted as dashed lines.
The dynamics of perturbations does not depend on the choice of initial conditions because the system is rapidly driven to the attractor solution corresponding to the growing mode of $\delta_c$. The different regimes explained in the analytical results of 
Sec.~\ref{Sec:growth} can be easily recognized. 
In particular, we observe the suppressed growth of $\delta_c$ for modes inside the effective DE sound horizon ($k>k_s$) with respect to the non-interacting case and that the two fluids comove 
($\theta_c \simeq \theta_d$) in the strong coupling regime.}
\label{Fig:numerical}
\end{figure}

In Fig.~\ref{Fig:numerical} the evolution of background and 
perturbed quantities is plotted for $b=-10^4\rho_0$ for $r=10^{-6}$, 
where $\rho_0=3H_0^2/(8\pi G)$ 
is today's critical density. 
In the upper panels, we show the relevant background quantities where we can see how the dark radiation energy density 
$\rho_{dr}=\rho_{\Lambda}ra^{-4}$ is negligible 
relative to $\rho_c$ and $\rho_{\Lambda}$.
We also observe the time at which the strong coupling regime ($|b|>\rho_c$) sets in. 
As we estimated in Eqs.~(\ref{ceff1}) and (\ref{ceff2}), 
the inverse of the effective DE horizon scale evolves as 
$k_s \propto \tau^3$ in the regime $\rho_c/|b| \gg 1$. 
After entering the strong coupling regime, $k_s$ approaches a constant 
during matter dominance and it starts to increase 
after the onset of cosmic acceleration. 
The scale dependence of perturbations arises by 
the fact that the epoch at which the DE sound 
horizon crossing occurs (or does not occur) 
depends on the wavenumber $k$.

In the middle and lower panels of Fig.~\ref{Fig:numerical}, 
we show the evolution of the density contrasts and velocity potentials 
for the modes $k=2\times 10^3 \cH_0$ and $k=2\times 10^2\cH_0$, respectively, 
where ${\cal H}_0$ is today's Hubble parameter.
Although these modes may correspond to the non-linear scales of structure formation, 
we have chosen for illustrative purposes to understand the behavior of CDM 
and DE perturbations in the presence of couplings.

For the model parameters under consideration, 
the mode $k=2\times 10^3 \cH_0$ has been always inside the 
DE sound horizon ($k>k_s$) by today. 
As we showed in Eq.~(\ref{delcNeq}), the CDM 
density contrast grows as $\delta_{c} \propto a$ in the weak coupling 
regime ($\rho_c \gg |b|$) of matter era. 
The DE density contrast first exhibits a rapid oscillation 
due to the dominance of the homogeneous mode $\delta_d^{\rm H}$ 
over the special solution $\delta_d^{\rm slow}$ 
in Eq.~(\ref{deldsum}). This fast oscillation ceases 
after $\delta_d^{\rm slow}$ dominates over $\delta_d^{\rm H}$, 
whose property can be seen in the middle left panel of
Fig.~\ref{Fig:numerical}. 
After the perturbations enter the strong coupling regime 
($\rho_c \ll |b|$), $\delta_c$ is nearly frozen 
as estimated by Eq.~(\ref{delcfro}).
This leads to the suppression for the growth of $\delta_c$ 
with respect to the non-interacting case 
(which is shown as dashed lines). 
In this regime, we can also confirm the relation 
(\ref{deldfro}), i.e., $\delta_d/(1+w_d) \simeq {\rm constant}$ 
with a suppressed amplitude relative to $\delta_c$.

In the middle right panel of Fig.~\ref{Fig:numerical}, we 
observe how the two fluids tend to move together, 
i.e., $\theta_c\simeq\theta_d$ in the strong coupling regime, 
which is again in accordance with our analytical estimation 
given in Eq.~(\ref{thetacd}).
This behaviour further illustrates the fact that the suppressed CDM clustering is related to the suppression of peculiar velocities induced by the DE dragging, whose pressure prevents the appearance of large peculiar motions. For the DE sector, it is apparent how the evolution of its perturbations, both the velocity and density, starts differing from the non-interacting 
evolution when the condition 
$\vert b\vert \simeq (1+w_d)\rho_d$ is met, 
while the CDM sector is not affected until the onset of 
the full strong coupling regime 
characterized by $\vert b\vert= \rho_c$.

For the mode $k=2\times 10^2\cH_0$ plotted in the bottom 
panels of Fig.~\ref{Fig:numerical}, the perturbations crossed 
outside the effective DE sound horizon ($k<k_s$) around the same 
epoch when they entered the strong coupling regime.
In the left panel, we can confirm that the DE and CDM density 
contrasts obey the adiabatic relation (\ref{adimode}) for $k<k_s$.
In this case the interacting term on the right 
hand-side of Eq.~(\ref{eq:deltacoscillator}) is negligible, 
so the evolution of $\delta_c$ is similar to that in the uncoupled case.
Since the asymptotic value of $w_d$ is $-1$, 
$\delta_d$ tends to be smaller than $\delta_c$ at late times 
due to the adiabatic relation $\delta_d=(1+w_d)\delta_c$. 
We note that, even though the growth of $\delta_c$ is not 
suppressed for $k<k_s$, the CDM and DE velocities approach 
a same value in the strong-coupling regime (see the right panel).
This fact is in agreement with the analytic estimation 
given in Sec.~\ref{velosec}.

\begin{figure}[!t]
\includegraphics[width=0.45\textwidth]{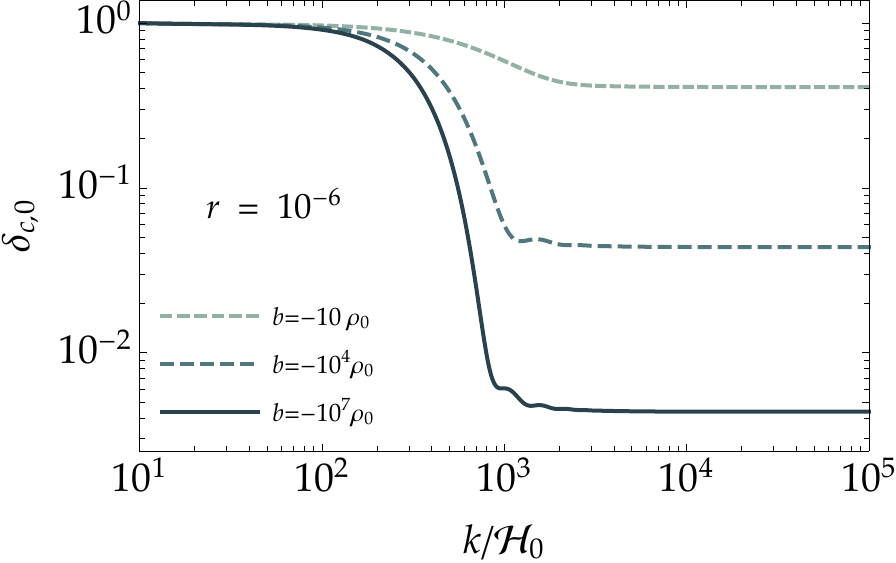}
\includegraphics[width=0.45\textwidth]{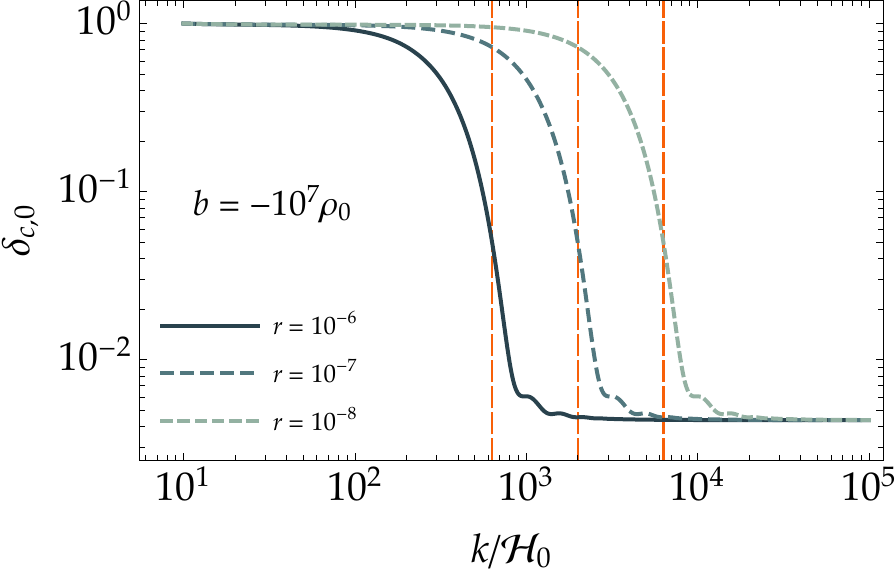}
\includegraphics[width=0.45\textwidth]{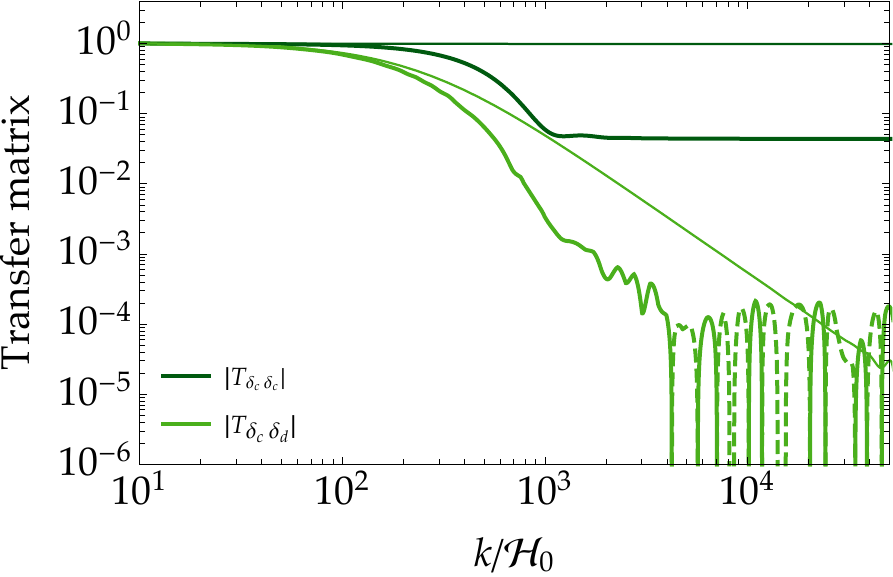}
\includegraphics[width=0.45\textwidth]{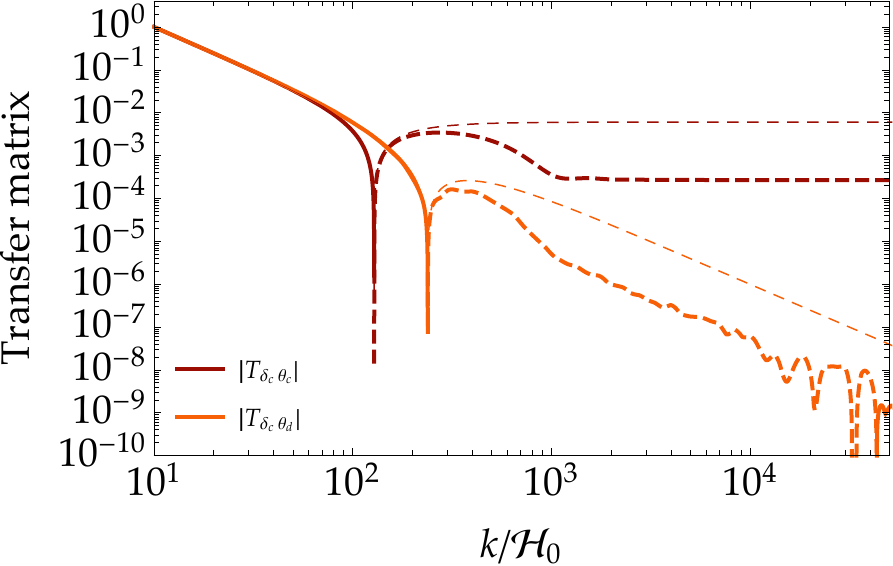}
\caption{In the {\bf upper panels}, we plot today's value of CDM 
density contrast $\delta_{c,0}$ 
versus $k/\cH_0$. As explained in the main text, this gives a good approximation to the transfer function 
for $\delta_c$. The left panel shows the dependence with 
respect to $b$ for $r=10^{-6}$, while the right panel shows 
how it varies with $r$ for $b=-10^7 \rho_0$.
We can see how the parameter $r$ mainly determines the values of
$k_s$ (shown by the vertical lines in the right panel) around 
which there is a suppression of $\delta_c$ and 
the parameter $b$ fixes the suppression. In these figures, 
we notice some small oscillations that are reminiscent of 
acoustic oscillations produced by dark radiation in the DE 
component as the modes cross the effective DE sound horizon. 
In the {\bf lower panels}, we present all the relevant elements of the transfer matrix for $\delta_c$ with $b=-10^{4}\rho_0$ and $r=10^{-6}$ involving density contrasts (lower left) and velocity potentials (lower right). We show in solid (dashed) lines the positive (negative) values of each matrix element. For comparison, we plot the non-interacting case in thinner lines to illustrate how the suppression affects all the transfer matrix components. These figures show how the diagonal term clearly dominates over the off-diagonal components, as claimed in the main text, 
which justifies neglecting them in the computation of the effect for the CDM density contrast.}
\label{fig:Tk}
\end{figure}

In the upper left panel of Fig.~\ref{fig:Tk}, we plot the 
CDM density contrast evaluated today (denoted as 
$\delta_{c,0}$) as a function of $k/\cH_0$ 
for $r=10^{-6}$ with three different values of $b$. 
Again, we present the results comprising the non-linear 
clustering regimes ($k \gtrsim 500\cH_0$) for illustrative purposes. 
As we discussed in Sec.~\ref{densec}, the growth of $\delta_c$ 
is suppressed in the strong coupling regime 
for small-scale modes inside the effective DE sound horizon. 
For increasing $|b|$, the perturbations enter the strong 
coupling regime earlier, so the modes with suppressed 
growth of $\delta_c$ span in the region with smaller values of $k$.
From Eq.~(\ref{rdelta}), the amplitude 
of CDM density contrast has the dependence 
$\delta_c/\delta_c^{b=0} \propto |b|^{-1/3}$, 
whose property can be confirmed in Fig.~\ref{fig:Tk}.

In the upper right panel of Fig.~\ref{fig:Tk}, we show 
$\delta_{c,0}$ versus $k/\cH_0$ for $b=-10^7 \rho_0$ with 
three different values of $r$. 
Since $k_s$ has the dependence (\ref{kses}),
the smaller $r$ leads to a shift of the region 
with suppressed values of $\delta_c$ toward larger $k$.
This means that we need to go to large values 
of $r$ to include larger scales 
in the suppression band. 
Since this parameter has an upper bound imposed by the maximum fraction of dark radiation in the early Universe, there should be an upper limit 
for the largest scale that can undergo a clustering suppression. 
As already mentioned, we need $r\lesssim 10^{-6}$ 
for the initial fraction of dark radiation to be smaller 
than 1\,\% in comparison to standard radiation.
In the strong coupling regime of matter dominance, we showed 
that $k_s$ is constant, see Eq.~(\ref{ceff2}).
On using the approximations $ra^{-4} \ll 1$ and 
$\rho_d \simeq \rho_{\Lambda}$ in this period, 
the effective DE sound speed is given by 
$c_{\rm eff} \simeq (2/3)\sqrt{\Omega_{\Lambda}/\Omega_c} 
\sqrt{r}\,a^{-1/2}$.
Since ${\cal H} \simeq H_0 \sqrt{\Omega_c}\,a^{-1/2}$ 
during the matter domination, it follows that 
\be
k_s \simeq
\frac{3}{2} \frac{\cH_0}{\sqrt{r}} 
\frac{\Omega_c}{\sqrt{\Omega_{\Lambda}}}\,.
\ee
For $\Omega_c \simeq 0.3$ and $\Omega_\Lambda \simeq 0.7$, 
the upper bound $r\lesssim 10^{-6}$ translates to the lower bound 
on $k_s$ with the minimum value 
$k_{s,{\rm min}}\sim 500 \cH_0$. 
As we observe in the upper right panel 
of Fig.~\ref{fig:Tk}, the transition of $\delta_{c,0}$ with respect to $k$
is not very sharp, so there are scales larger than this bound 
(say, $100\cH_0 \lesssim k \lesssim 500 \cH_0$) where the CDM 
density contrast is subject to suppression.

Let us also discuss the off-diagonal terms for 
the transfer matrix of perturbations expressed as 
a vector form $\vec{X}=(\delta_c,\delta_d,\theta_c,\theta_d)$. 
Denoting $\vec{X}_{\rm ini}$ and $\vec{X}_0$ as the initial and 
present values of $\vec{X}$, respectively, the transfer matrix 
$\hat{T}$ relates them according to 
$\vec{X}_0=\hat{T} \vec{X}_{\rm ini}$. 
In particular, for the CDM density 
contrast, 
we have
\be
\delta_{c,0}=T_{\delta_c\delta_c}\delta_{c,\rm ini}+T_{\delta_c\delta_d}\delta_{d,\rm ini}+T_{\delta_c\theta_c}\theta_{c,\rm ini}+T_{\delta_c\theta_d}\theta_{d,\rm ini}\,.
\ee
Numerically, the components of the transfer matrix 
relevant to $\delta_c$ can be computed by evaluating $\delta_c$ at the final time with initial conditions given by the vectors of the canonical basis, i.e., with $\vec{X}_{\rm ini}=(1,0,0,0)$, $\vec{X}_{\rm ini}=(0,1,0,0)$, $\vec{X}_{\rm ini}=(0,0,1,0)$, $\vec{X}_{\rm ini}=(0,0,0,1)$, respectively. 
We have computed all the relevant components of 
the transfer matrix in Fig.~\ref{fig:Tk}, where we observe 
that the diagonal component $T_{\delta_c\delta_c}$ is clearly the dominant one 
over the others. 
This together with the fact that the CDM perturbations have a larger amplitude  at the onset of the interacting regime, justifies the initial conditions we have chosen to study 
the suppressed clustering.

\begin{figure}[!t]
\includegraphics[width=0.49\textwidth]{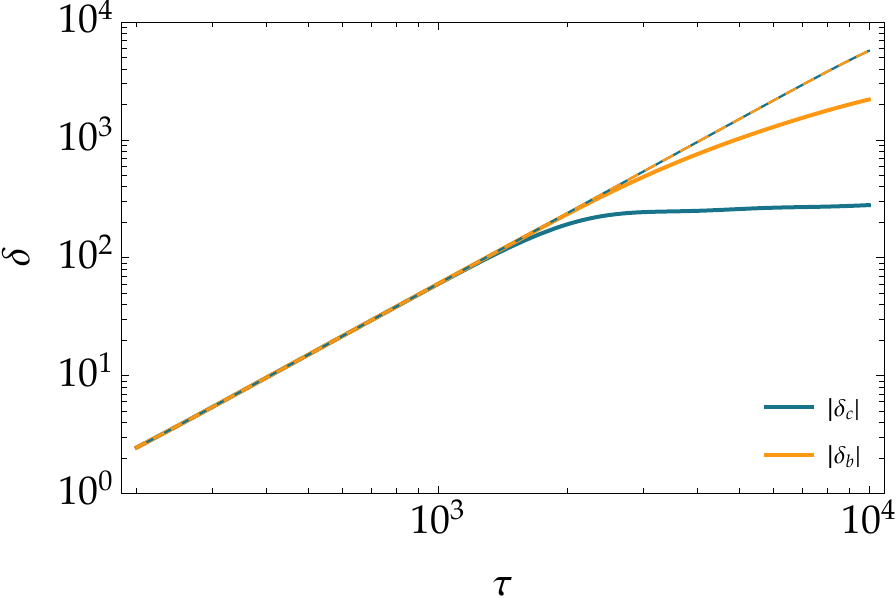}
\includegraphics[width=0.49\textwidth]{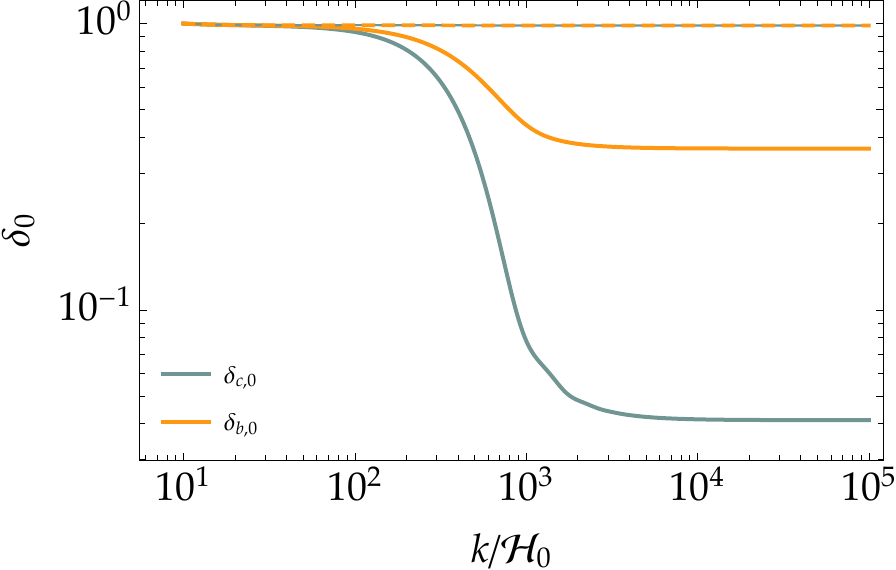}
\caption{In this figure, we illustrate the effect of CDM and DE momentum 
exchange on the baryon density contrast $\delta_b$ for $b=-10^{4}\rho_0$ and $r=10^{-6}$. The left panel shows the evolution of $\delta_c$ and 
$\delta_b$ for a Fourier mode $k=2\times 10^{3}\cH_0$ (which is 
inside the effective DE sound horizon). Unlike $\delta_c$, we see how $\delta_b$ is subject to a much milder suppression as compared to the non-interacting case (thinner lines). 
This property is also confirmed in the right panel, where today's values of 
$\delta_b$ and $\delta_c$ versus $k/\cH_0$ are shown.
This milder suppression arises because the effect on the baryon perturbation 
is only indirect due to the less clustering of CDM that induces a smaller gravitational potential so baryons fall into shallower wells and, therefore, they cluster less. On the other hand, the CDM is affected by the pressure of the DE component that prevents the clustering in a more direct and efficient manner.}
\label{fig:withbaryons}
\end{figure}

Finally, we also solved the perturbation equations of baryons
and found that, unlike $\delta_c$, the growth of $\delta_b$ is more mildly suppressed in the regime $\rho_c \ll |b|$ (see Fig.~\ref{fig:withbaryons}). 
The underlying reason is that, while CDM is directly affected by the interaction so that the DE pressure prevents the clustering, the baryons only feel the effect of the reduced clustering of CDM through the  smaller gravitational potential that gives rise to a weaker clustering as compared to the non-interacting case. 

\section{Conclusions}

In this work, we have explored a scenario where the dark sector of 
the Universe contains CDM and DE described by perfect fluids with the Schutz-Sorkin action that 
interact via a velocity-dependent coupling. The interaction is 
characterized by the function 
$f(Z)$, where $Z=g_{\mu \nu}u_{c}^{\mu}u_{c}^{\nu}$ is the scalar product 
of four velocities. 

In many phenomenological approaches taken in the literature, 
the interactions in the dark sector are added 
by hand at the background level. 
A  drawback of introducing the interactions at the background level 
is that the study of the perturbations (which is of paramount importance for testing the theoretical and phenomenological viability of the models) requires a covariantization of the interaction and this process inevitably comes in with ambiguities. Our scenario naturally avoids this problem 
because the background and perturbation
equations of motion unambiguously 
follow from an explicit action of perfect fluids with a momentum exchange. 
We also note that the interacting theory of Ref.~\cite{Asghari:2019qld}, 
that also avoids ambiguities by starting with a covariant formulation, is 
different from ours in that the former introduced a velocity-dependent 
coupling at the level of the continuity equations. 

Due to the nature of the interaction, the only modification to the background equations appears as a constant term $f(Z)$ with $Z=-1$. 
Since this term can be absorbed into a cosmological constant, 
the momentum exchange does not modify the dynamical evolution of the background cosmology. However, the interaction affects the perturbation 
equations of the CDM and DE velocity potentials through 
the momentum exchange.
We have derived the linear perturbation equations of motion without 
fixing gauges and obtained the conditions for the absence of ghosts and Laplacian instabilities. These stability conditions can be easily guaranteed by imposing the usual weak/null energy conditions and a negative coupling constant $b$ in the dark sector. The fact that the interaction only affects the Euler equations has important implications from phenomenological and observational viewpoints. Firstly, the background is oblivious to the interaction and, consequently, the homogeneous evolution cannot constrain the corresponding coupling parameter. Secondly, the perturbed continuity equation remains the same as in $\Lambda$CDM so the relation between the density field and the divergence of the velocity field still holds even though the evolution of both is modified. 

After developing the general formalism of dealing with cosmological perturbations 
in our interacting theory, we proposed a concrete model 
in which the DE sector contains a cosmological constant and dark radiation. 
In this model the DE fluid behaves as dark radiation with the equation of state $w_d \simeq 1/3$ at early times, 
so this allows a possibility for alleviating the $H_0$ tension present in the $\Lambda$CDM model.
After the perturbations enter the strong coupling regime characterized by $|b|>\rho_c$, the peculiar velocity of 
CDM approaches that of DE, 
i.e., $\theta_c \simeq \theta_d$. 
For the wavenumber $k$ in the range $k>k_s$, where $k_s={\cal H}/c_{\rm eff}$ 
is the inverse of an effective DE sound horizon associated with the 
propagation speed squared (\ref{ceff}), we have analytically shown that the CDM density contrast $\delta_c$ approaches a constant 
in the strong coupling regime of matter dominance. 
This results in the suppression for the growth of $\delta_c$ 
in comparison to the uncoupled case ($b=0$).
For the modes $k<k_s$, the density contrasts in the region $|b|>\rho_c$ evolve adiabatically ($\delta_d \simeq (1+w_d) \delta_c$), without the suppressed growth 
of $\delta_c$.
We have corroborated our analytical findings by numerically solving the perturbation equations and found perfect agreement.

The suppression of the CDM density contrast 
found in this work is in line with previous findings in the literature supporting the idea that the momentum exchange in the dark sector can alleviate the $\sigma_8$ 
tension. Moreover, in our concrete interacting model, there exists 
dark radiation in the early Universe that may ease 
the $H_0$ tension.
These properties encourage further investigations on their cosmological viability.
For the scenario considered in this work, it would be desirable to perform a detailed fit to cosmological data to confirm its ability to resolve said tensions. Work is in progress in this direction.

\section*{Acknowledgements}

 JBJ, DB, DF and FATP acknowledge support from the {\it Atracci\'on del Talento Cient\'ifico en Salamanca} programme, from project PGC2018-096038-B-I00 by {\it  Spanish Ministerio de Ciencia, Innovaci\'on y Universidades} and {\it Ayudas del Programa XIII} by USAL. DF acknowledges support from the programme {\it Ayudas para Financiar la Contratación Predoctoral de Personal Investigador (ORDEN EDU/601/2020)} funded by Junta de Castilla y Leon and European Social Fund. 
ST is supported by the Grant-in-Aid for Scientific Research Fund of the JSPS No.\,19K03854.

\appendix
\section{Equations in synchronous gauge}

In this Appendix, we give the perturbation equations in the synchronous gauge defined by the perturbed line element 
\begin{equation}
{\rm d} s^2=a^2(\tau) \left[-{\rm d} \tau^2 + (\delta_{ij}+h_{ij}) {\rm d} x^i {\rm d} x^j \right]\,,
\end{equation}
where the perturbed spatial metric components 
are written in terms of the scalar perturbations $h$ and $\eta$ as 
$h_{ij} = {\rm diag}(-2\eta, -2\eta, h+\eta)$. 
In this gauge, the continuity and Euler equations 
for CDM and DE perturbations are given by 
\begin{eqnarray}
\left(\delta_{c}^{\rm sync}\right)'&=&-\left(\theta_{c}^{\rm sync} + \frac{1}{2} h'\right)\,,\\
\left(\theta_{c}^{\rm sync}\right)'&=&-\mathcal{H} \theta_{c}^{\rm sync}  
+b\frac{3 {\cal H} (1+w_d) \rho_d 
[\theta_c^{\rm sync}-(1+c_d^2) \theta_{d}^{\rm sync}]
-k^2  c_d^2 \rho_d \delta_{d}^{\rm sync}}
{(1+w_d)\rho_d(\rho_c-b)-b\rho_c}\,,\\
\left(\delta_{d}^{\rm sync}\right)'& = & -3 \mathcal{H} (c_d^2-w_d)\delta_{d}^{\rm sync} -(1+w_d)\left(\theta_{d}^{\rm sync} + \frac{1}{2}h'\right)\,, \\
\left(\theta_{d}^{\rm sync}\right)'&=&(-1+3c_d^2) \mathcal{H} \theta_{d}^{\rm sync} 
+\frac{\rho_c[k^2 c_d^2 \rho_d \delta_d^{\rm sync}+
3 {\cal H} b\{(1+c_d^2) \theta_d^{\rm sync}-\theta_c^{\rm sync}\}]
-k^2 b c_d^2 \rho_d \delta_d^{\rm sync}}
{(1+w_d)\rho_d(\rho_c-b)-b\rho_c}\,.
\end{eqnarray}
The suppression of the CDM density contrast explained in detail in the Newtonian gauge also occurs in 
the same manner as in the synchronous gauge, since the density contrast for modes deep inside the horizon is gauge-invariant to a good approximation.

\bibliography{bibinteractingfluids}

\end{document}